\documentclass[%
 aip,
 amsmath,amssymb,
 reprint,%
]{revtex4-1}

\usepackage{graphicx}
\usepackage{dcolumn}
\usepackage{bm}

\usepackage[utf8]{inputenc}
\usepackage[T1]{fontenc}
\usepackage{mathptmx}
\usepackage{etoolbox}
\usepackage{mathrsfs}
\usepackage{xcolor}
\makeatletter
\def\@email#1#2{%
 \endgroup
 \patchcmd{\titleblock@produce}
  {\frontmatter@RRAPformat}
  {\frontmatter@RRAPformat{\produce@RRAP{*#1\href{mailto:#2}{#2}}}\frontmatter@RRAPformat}
  {}{}
}%
\makeatother
\begin{document}

\preprint{AIP/123-QED}

\title[]{Tuning photoacoustics with nanotransducers via Thermal Boundary Resistance and Laser Pulse Duration}
\author{Michele Diego}
\email{michele.diego@univ-lyon1.fr}
\affiliation{FemtoNanoOptics group, Universit\'e de Lyon, CNRS, Universit\'e Claude Bernard Lyon1, Institut Lumi\`ere Mati\`ere, F-69622 Villeurbanne, France
}%

\author{Marco Gandolfi}
\affiliation{CNR-INO, via Branze 45, 25123 Brescia, Italy}
\affiliation{Department of Information Engineering, Universit\`a di Brescia, via Branze 38, 25123 Brescia, Italy}
\affiliation{Interdisciplinary Laboratories for Advanced Materials Physics (I-LAMP) and Dipartimento di Matematica e Fisica, Universit\`a Cattolica del Sacro Cuore, via della Garzetta 48, Brescia, I-25133, Italy}

\author{Stefano Giordano}
\affiliation{Universit\'e de Lille, CNRS, Centrale Lille, Universit\'e Polytechnique Hauts-de-France, UMR 8520 — IEMN — Institut d’\'Electronique de Micro\'electronique et de Nanotechnologie, F-59000 Lille, France}

\author{Fabien Vialla}
\affiliation{FemtoNanoOptics group, Universit\'e de Lyon, CNRS, Universit\'e Claude Bernard Lyon1, Institut Lumi\`ere Mati\`ere, F-69622 Villeurbanne, France}

\author{Aur\'elien Crut}
\affiliation{FemtoNanoOptics group, Universit\'e de Lyon, CNRS, Universit\'e Claude Bernard Lyon1, Institut Lumi\`ere Mati\`ere, F-69622 Villeurbanne, France}

\author{Fabrice Vall\'ee}
\affiliation{FemtoNanoOptics group, Universit\'e de Lyon, CNRS, Universit\'e Claude Bernard Lyon1, Institut Lumi\`ere Mati\`ere, F-69622 Villeurbanne, France}

\author{Paolo Maioli}
\affiliation{FemtoNanoOptics group, Universit\'e de Lyon, CNRS, Universit\'e Claude Bernard Lyon1, Institut Lumi\`ere Mati\`ere, F-69622 Villeurbanne, France}

\author{Natalia Del Fatti}
\affiliation{FemtoNanoOptics group, Universit\'e de Lyon, CNRS, Universit\'e Claude Bernard Lyon1, Institut Lumi\`ere Mati\`ere, F-69622 Villeurbanne, France}
\affiliation{Institut Universitaire de France (IUF), F-75005, Paris, France}

\author{Francesco Banfi}
\affiliation{FemtoNanoOptics group, Universit\'e de Lyon, CNRS, Universit\'e Claude Bernard Lyon1, Institut Lumi\`ere Mati\`ere, F-69622 Villeurbanne, France}

\date{\today}

\begin{abstract}
\vspace{0.05cm}
Copyright 2024 M. Diego, M. Gandolfi, S. Giordano, F. Vialla, A. Crut, F. Vallée, P.
Maioli, N. Del Fatti and F. Banfi. This article is distributed under a Creative Commons Attribution-NonCommercial 4.0 International (CC BY-NC) License. This article appeared in \textit{Appl. Phys. Lett.} 121, 252201 (2022) and may be found at https://doi.org/10.1063/5.0135147\\
\vspace{0.05cm}
\\
The photoacoustic effect in liquids, generated by metal nanoparticles excited with short laser pulses, offers high contrast imaging and promising medical treatment techniques. Understanding the role of the thermal boundary resistance (TBR) and the laser pulse duration in the generation mechanism of acoustic waves is essential to implement efficient photoacoustic nanotransducers. This work theoretically investigates, for the paradigmatic case of water-immersed gold nanocylinders, the role of the TBR and of laser pulse duration in the competition between the launching mechanisms: the thermophone and the mechanophone. In the thermophone, the nanoparticle acts as a nanoheater and the wave is launched by water thermal expansion. In the mechanophone, the nanoparticle directly acts as a nanopiston.
Specifically, for a gold-water interface, the thermophone prevails under ns light pulse irradiation, while the mechanophone dominates shortening the pulse to the 10 ps regime. For a graphene-functionalized gold-water interface, instead, the mechanophone dominates over the entire range of explored laser pulse durations. Results point to high-TBR, liquid-immersed nanoparticles as potentially efficient photoacoustic nanogenerators, with the advantage of keeping the liquid environment temperature unaltered.
\end{abstract}

\maketitle

Nanoscale photoacoustics generation in liquids, owing to its potential in nanoimaging and therapeutic applications, is a flourishing topic at the cross-road of condensed matter physics, nanomedicine and material science\cite{emelianov2009photoacoustics, li2009photoacoustic,xu2006photoacoustic, wang2017photoacoustic, li2015gold}. In this context, liquid-immersed metal nanoparticles have proven to be efficient photoacoustic generators due to their tunable optical absorption properties \cite{yang2009nanoparticles, mantri2020engineering, el2008shape, garcia2020optimizing}, high contrast imaging features \cite{luke2012biomedical, pan2011molecular} and biocompatibility \cite{shukla2005biocompatibility, craciun2017surface}. Great efforts have been devoted to optimise the parameters allowing a more efficient photoacoustic conversion, such as size, geometry \cite{garcia2020optimizing, pramanik2016simulating, guiraud2021thermoacoustic, ronchi2021discrimination} and transducer materials\cite{lee2018efficient}. Yet, despite its relevance for applications, the combined effects of the pulse temporal width, $\tau_{L}$, \cite{gandolfi2020optical, prost2015photoacoustic, diego2022Ultrafast} and the thermal boundary resistance\cite{chen2011silica, repenko2018strong,cavigli2020impact} (TBR) tunabilities remain relatively unexplored and lack a thorough rationalization.\\
\begin{figure}[t]
\centering
\includegraphics[width=0.45\textwidth]{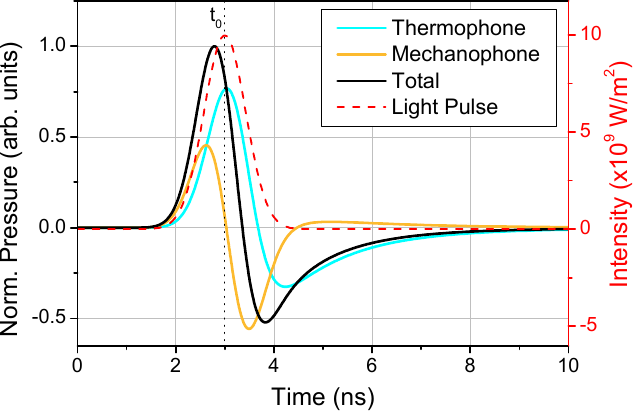}
\caption{
Left axis: pressure time evolution in water at $r=R_{gnc}+5$ nm for $\tau_L = 1$ ns, $\mathscr{R}=1 \times 10^{-7}$ m$^2$K/W. The curves are normalized to the total pressure maximum. Right axis: $I(t)$ (red dashed curve).}
\label{fig:pressure1ns}
\end{figure}
In brief, the photoacoustic effect of an individual liquid-immersed metal nanoparticle consists of three steps: (i) absorption of the laser pulse by the nanoparticle and its temperature rise, (ii) thermal interaction between the nanoparticle and the liquid environment and (iii) generation of the acoustic wave in liquid. The acoustic wave in liquid is triggered by two launching mechanisms: the mechanophone and the thermophone effects. The former is due to the thermal expansion of the metal nanoparticle, the latter mechanism to that of the liquid environment with the nanoparticle acting as a nanoheater.
The photoacoustic generation in these systems is typically investigated under ns laser pulses irradiation, in which case the mechanophone contribution is, in most instances, disregarded \cite{chen2012environment, shahbazi2019photoacoustics, gandolfi2020optical, guiraud2019multilayer, guiraud2019two}.
Nevertheless, the mechanophone effect cannot be neglected in general, as recently demonstrated for the case of carbon nanotubes immersed in water\cite{diego2022Ultrafast}.
\begin{figure*}[t]
\centering
\includegraphics[scale=0.65]{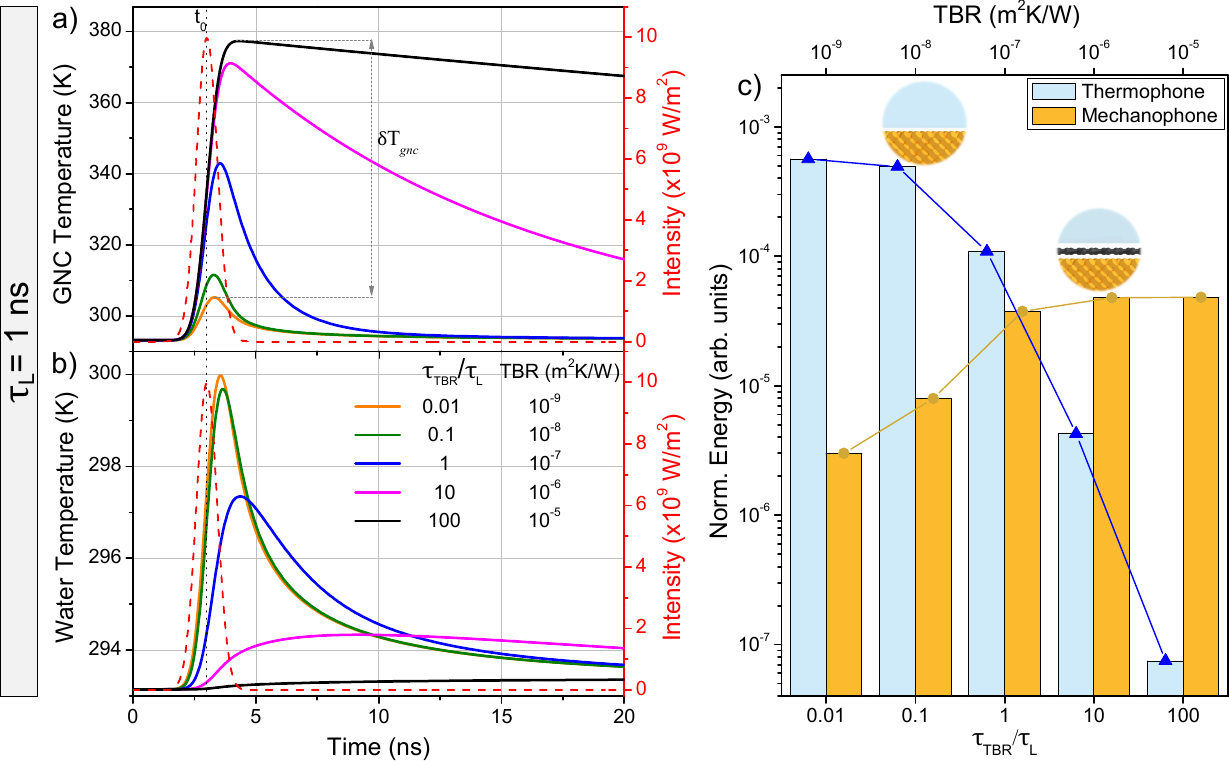}
\caption{Laser pulse: $\tau_L=1$ ns. Panels (a,b): temperature time evolution, for increasing values of $\tau_{TBR}/\tau_L$ and the corresponding TBR values, in the GNC ($r=0$ nm): panel (a); in proximal water ($r=R_{gnc}+5$ nm): panel (b). Right axis: $I(t)$ (red dashed line) maximized at the time $t_{0}$. $\delta T_{gnc}$: difference between the GNC peak temperatures between the cases of $\tau_{TBR}/\tau_{L}\sim$ 100 and 0.01. Panel (c): normalized mechanical energy generated in water by the thermophone and the mechanophone effects for different $\tau_{TBR}/\tau_L$ (bottom axis) and the corresponding TBRs (top axis). The ratios $\tau_{TBR}/\tau_{L}$ are rounded to the first significant figure. Values for Au/water and graphene-functionalized Au/water interface are identified by the two round sketches.}
\label{fig:histo1ns}
\end{figure*}
Once the size and composition of the nanoparticles and of the surrounding liquid are fixed by the specific application constraints, the relative contributions of the thermophone vs mechanophone effects may be tuned upon varying the TBR at the nanoparticle/liquid interface (interface engineering) and $\tau_{L}$, their interplay making the focus of the present work.\\
We theoretically investigate the role of the thermophone and mechanophone effects in acoustic wave generation for the paradigmatic case of a water-immersed gold nanocylinder (GNC) of radius $R_{gnc}$=10 nm and of high aspect ratio, because of their relevance in bio-medical applications \cite{he2020application, manohar2011gold, cavigli2014size, cavigli2017photo, chen2019miniature, dhada2020gold, knights2019optimising}. Formally, the GNC is assumed infinitely extended along its axis, the problem thus being radially symmetric with $r$ the radial coordinate. 
The model, detailed in SI, comprises three steps: optics, thermics and mechanics.
As for the optics, the system, at equilibrium at $t=0$ s, is excited with a laser pulse at 530 nm wavelength, i.e. within water transparency window. The light intensity (W/m$^2$) has a Gaussian intensity  profile,
$I(t) = 2\sqrt{\frac{\ln{(2)}}{\pi}}  \frac{\Phi}{\tau_L} \exp{\big[-4\ln{(2)}\frac{(t-t_0)^2}{\tau_L^2}\big]}$,
where $\Phi=$ 10 J/m$^2$ is the pulse fluence, which is kept constant for all $\tau_L$ values, with $\tau_L$=1 ns, 100 ps, 10 ps its temporal full-width-half-maximum, while the pulse temporal peak occurs at $t_0=3 \tau_L$. Via the GNC absorption cross section we then calculate the absorbed power density.
The latter serves as the source term for the thermics. The temperature $T(r,t)$ throughout the system (both GNC and water) is then obtained solving the thermal diffusion equation while imposing the continuity of the heat flux at the GNC/water interface,
$\Vec{q}=\frac{1}{\mathscr{R}}[T(R_{gnc}-,t)-T(R_{gnc}+,t)] \hat{r}$, which is controlled by the TBR, $\mathscr{R}$, with $T(R_{gnc}\pm, t)$ the temperature at the inner (-) and outer (+) side of the interface and $\hat{r}$ its normal vector.
$T(r,t)$ serves as the source terms for the mechanics via the thermal expansion coefficients of both the GNC and water, ultimately yielding the pressure, $p(r,t)$, and the radial velocity field, $v_{r}(r,t)$ in water. With $p$ and $v_{r}$ at hand, the acoustic Poynting vector, $\textbf{P}$ (W/m$^2$), and from it the mechanical energy radiated in water, $U$, are retrieved. The thermophone and mechanophone contribution to the total $p(r,t)$ and $U$ are calculated forcing to zero the GNC and water thermal expansion coefficients, respectively.\\
\indent The first conclusion that can be drawn from simulations results is that the mechanophone effect needs to be accounted for.
Fig.\ref{fig:pressure1ns} shows $p(t)$ in water, 5 nm from the GNC/water interface, together with the thermophone and mechanophone contributions. 
Results are for the case of $\tau_L = 1$ ns and $\mathscr{R}=1 \times 10^{-7}$ m$^2$K/W. The latter is representative of the general cases that might be encountered: its order of magnitude falls between that of the  Au/water, $1 \times 10^{-8}$ m$^2$K/W\cite{plech2004laser,stoll2015time,gandolfi2020optical,wilson2022curvature}, and that of the graphene-functionalized Au/water interface\cite{herrero2022ultra}, $1 \times 10^{-6}$ m$^2$K/W.
The thermophone and the mechanophone contributions have similar amplitudes, thus both contributing to the total pressure signal. 
\begin{figure*}[t]
\centering
\includegraphics[scale=0.65]{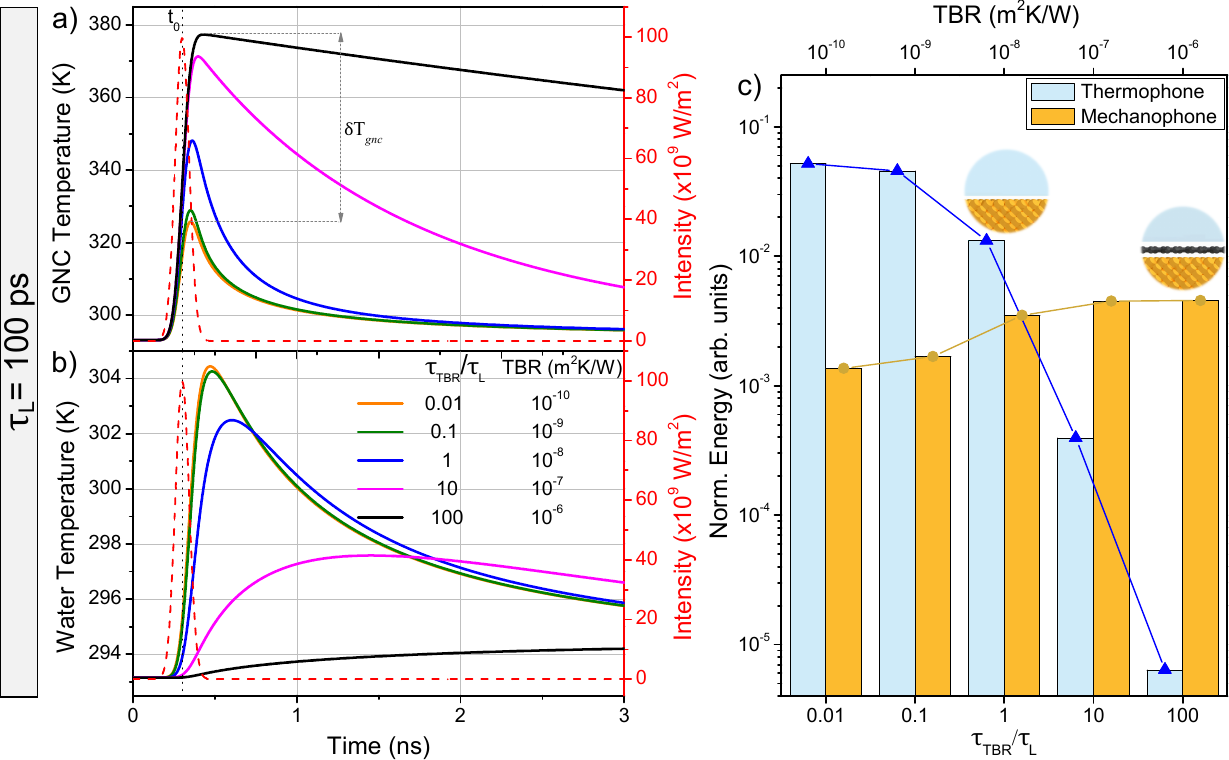}
\caption{Laser pulse: $\tau_L=100$ ps. Same caption as for Fig.\ref{fig:histo1ns}. The TBR values are varied so as to span the same values of $\tau_{TBR}/\tau_L$ as for the 1 ns pulse case.}
\label{fig:histo100ps}
\end{figure*}
\\
\indent We now address the role played by the TBR and $\tau_{L}$ in the relative contribution between these two launching mechanisms.
On the thermal side, upon absorption of the laser pulse, the GNC raises its temperature on a time scale $\tau_{L}$. It cools down on a time scale $\tau_{th}$ transferring heat to the proximal water and raising the temperature of the latter. Finally, the GNC and the proximal water diffuse heat to distant water, relaxing to the initial temperature.
The timescale $\tau_{th}$ has contributions from the TBR and heat diffusion effects arising from the GNC and proximal water thermal impedances
\footnote{Due to the intricate relaxation  dynamics, $\tau_{th}$ escapes a formal definition. A commonly adopted operative approach is to define it as the time necessary for the GNC temperature increase, triggered by a $\delta$-like excitation source, to fall to 1/e of its maximum\cite{diego2022Ultrafast}.}. We now discuss what might be intuitively foreseen in the two extreme-case scenarios.\\
For \textit{$\tau_{th}\gg\tau_{L}$}, energy from the laser pulse is delivered to the GNC on a time scale $\tau_{L}$, and, only after a time $\sim\tau_{th}$, the GNC temperature decreases substantially, while delivering heat to the proximal water. That is, on a time scale $\tau_{th}$ we should expect a high-temperature GNC, thermally isolated from the surrounding water still at its ambient temperature. On the mechanics side, the thermal expansion of the GNC should be at its maximum. On the contrary, the contribution of water thermal expansion should be at a minimum and set in on a timescale exceeding $\tau_{th}$. The relevance of the mechanophone effect with respect to the thermophone should thus be highest for cases in which $\tau_{th}\gg\tau_{L}$.\\
For \textit{$\tau_{th}\ll\tau_{L}$} the situation is the opposite. The laser feeds energy to the GNC on a time scale $\tau_{L}$, whereas the GNC delivers energy to the proximal water on a much faster time scale, $\tau_{th}$: the GNC absorbs energy from the laser pulse and concomitantly delivers it to the proximal water.  
In this scenario the peak GNC temperature should be at its minimum, whereas the proximal water temperature should reach its greatest value. Accordingly, on the mechanics side, the peak thermal expansion of the GNC should be at its minimum, and that of proximal water at its maximum. The relevance of the mechanophone effect with respect to the thermophone should thus be lowest for $\tau_{th}\ll\tau_{L}$.\\
\indent In first instance, the ratio $\tau_{th}/\tau_{L}$ therefore appears as a meaningful metric to inspect the thermophone to mechanophone transition. The TBR is though the \textit{only} material parameter that can be tuned\cite{han2017thermal, pham2013pressure, barisik2014temperature, vera2015temperature, banfi2012temperature, wu2016thermal, caplan2014analytical, ramos2016solid, kim2008molecular, vo2015nano, barrat2003kapitza, hu2012effect, wang2011role, hasan2019manipulating, yenigun2019effect, cao2018enhanced, tian2015enhancing}, the thermal properties of the GNC and water being fixed. Under a practical stand-point it is therefore desirable to parameterize the problem in terms of a thermal decay time linked to the TBR only, rather than to $\tau_{th}$, which comprises also the effect of proximal water and GNC thermal impedances. To this end, we link the TBR to the thermal decay time through the expression $\tau_{TBR}$=$\mathscr{R} R_{gnc} c_p \rho /2$, with $c_p$ and $\rho$ the Au specific heat and mass density, respectively. This relation is somewhat approximate\footnote{Formally, the GNC cooling is mono-exponential with a decay given by this formula, only for a Biot number Bi$\ll$1 and for isothermal water \cite{banfi2010ab}, i.e. the cooling is limited by the interfacial heat transfer only.}, nevertheless, it has the merit of providing a rule-of-thumb estimate.
In the following, we therefore parameterize simulations results in terms of $\tau_{TBR}/\tau_{L}$ rather than $\tau_{th}/\tau_{L}$.
We now inspect what happens varying the TBR for a fixed laser pulse duration.\\
\indent\textit{Nanosecond regime}.
Fig.\ref{fig:histo1ns} shows the GNC (panel (a)) and the proximal water (panel (b)) temperature dynamics. The curves are calculated for a fixed value of $\tau_{L}$=1 ns while varying the TBR, from $10^{-5}$ to $10^{-9}$ m$^2$K/W, so as to cover the range of $\tau_{TBR}/\tau_{L}$ from $10^{-2}$ to $10^{2}$.
For increasing $\tau_{TBR}/\tau_{L}$ the GNC maximum temperature, $\textit{max}\{T_{gnc}(t)\}$, increases from 305 K to 377 K, whereas that of proximal water, $\textit{max}\{T_{w}(t)\}$, decreases from 300 K to 293 K.\\
\begin{figure*}[t]
\centering
\includegraphics[scale=0.65]{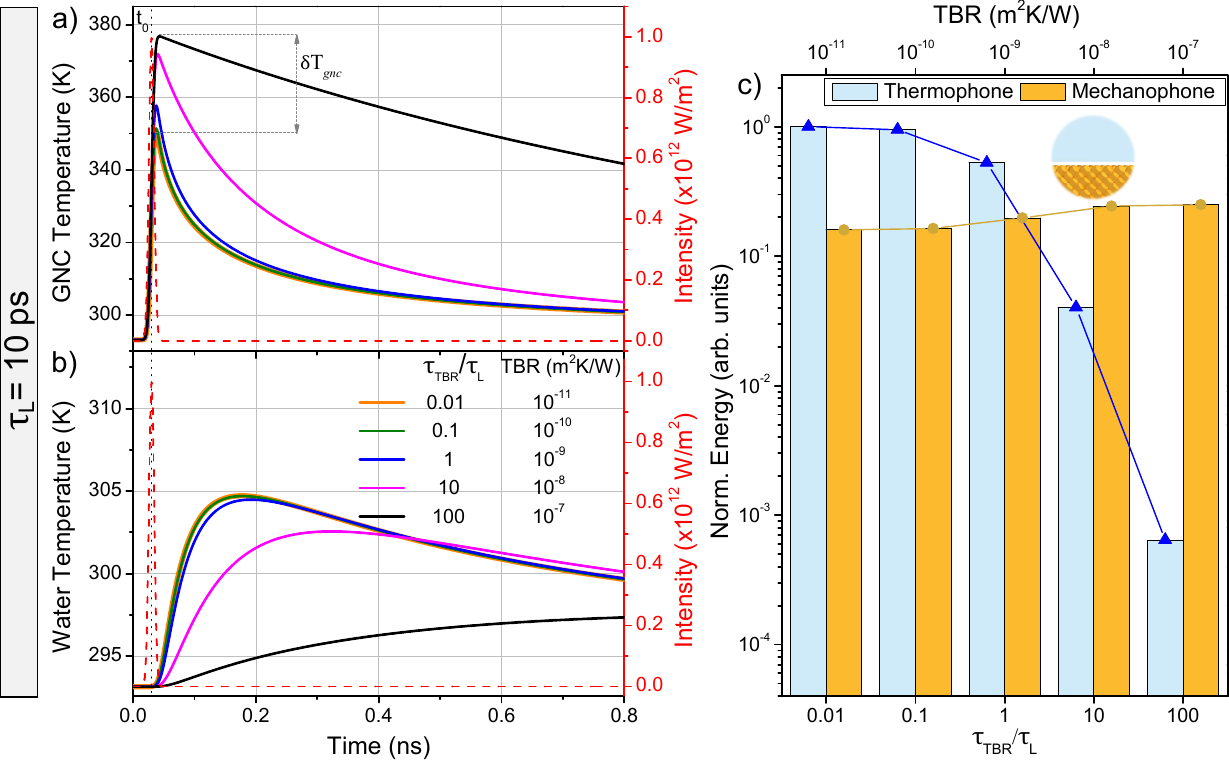}
\caption{Laser pulse: $\tau_L=10$ ps. Same caption as for Fig.\ref{fig:histo1ns}. The TBR values are varied so as to span the same values of $\tau_{TBR}/\tau_L$ as for the 1 ns pulse case. In panel (c) the case of the graphene-functionalized Au/water interface is here dominated by the mechanophone effect and is hence not reported.}
\label{fig:histo10ps}
\end{figure*}
The implications of the thermal problem on the competition between the thermophone and mechanophone contributions are shown in panel (c). The histogram shows the mechanical energy radiated in water by the sole thermophone (azure) and sole mechanophone (mustard) effects as a function of $\tau_{TBR}/\tau_{L}$ (bottom axis) and TBR (top axis). Energies are normalized to the maximum mechanical energy observed in water in our simulations (i.e. the thermophone contribution of the $\tau_L=10$ ps, $\tau_{TBR}/\tau_{L}=10^{-11}$ case that will be described later).
For increasing values of the TBR, acoustic wave generation in water switches from thermophone-dominated for $\tau_{TBR}/\tau_{L}\lesssim$ 1, 
to mechanophone-dominated for $\tau_{TBR}/\tau_{L}\gg$ 1, $\tau_{TBR}/\tau_{L}\approx$ 1 being a cross-over value between the two regimes.\\
\indent In real case scenarios, the TBR can be tuned engineering the GNC/water interface, and the acoustic wave launching mechanism accordingly, switching for instance from the thermophone for the bare Au/water interface, to the mechanophone for the graphene-functionalized Au/water interface, which cases are indicated by the two inset sketches.
\indent So far we spanned $\tau_{TBR}/\tau_{L}$ for a fixed value of $\tau_{L}$ while varying the TBR. The question then arises as to whether a similar physics holds true also for shorter laser pulses, thus eventually allowing for an additional knob, $\tau_{L}$, to select the launching mechanism.\\
\indent \textit{Picosecond regime}.
Fig.\ref{fig:histo100ps} and Fig.\ref{fig:histo10ps} are analogous to Fig.\ref{fig:histo1ns} but for the cases of $\tau_{L}$=100 ps and 10 ps, respectively. For sake of comparison, the TBR is now varied from $10^{-6}$ to $10^{-10}$ m$^2$K/W and from $10^{-7}$ to $10^{-11}$ m$^2$K/W for the case $\tau_{L}$=100 ps and 10 ps, respectively. These changes allow covering the same range of $\tau_{TBR}/\tau_{L}$ as for the case of the 1 ns laser pulse.
On the thermal side, Fig.\ref{fig:histo100ps}(a,b) and Fig.\ref{fig:histo10ps}(a,b) encompass the general trend observed in Fig.\ref{fig:histo1ns}(a,b): for increasing $\tau_{TBR}/\tau_{L}$, $\textit{max}\{T_{gnc}(t)\}$ increases whereas $\textit{max}\{T_{w}(t)\}$ decreases. While reducing $\tau_{L}$, a relevant parameter for the following discussion is $\delta T_{gnc}$, defined as
the difference between the GNC peak temperatures between the cases of $\tau_{TBR}/\tau_{L}$=10$^2$ and 10$^{-2}$.
$\delta T_{gnc}$ ranges from 72 K for $\tau_{L}$=1 ns, to 27 K for $\tau_{L}$=10 ps, because of the increase of the peak temperature for the case of $\tau_{TBR}/\tau_{L}$=$10^{-2}$ while transitioning from Fig.\ref{fig:histo1ns}(a), across Fig.\ref{fig:histo100ps}(a) to Fig.\ref{fig:histo10ps}(a). 
Among the physical reasons behind this trend, is that reducing $\tau_{L}$ for a fixed value of $\tau_{TBR}/\tau_{L}$=$10^{-2}$ implies reducing the TBR, eventually to a point where the interfacial heat transfer is no more the limiting process, the GNC and the proximal water thermal impedances remaining as the only factors controlling the thermal dynamics\footnote{As an extreme case scenario, if we were to nullify the TBR we would have $\tau_{TBR}$=0. In such a situation, the GNC thermal dynamics is entirely ruled by the GNC and proximal water thermal inertia.}.\\
The implications of the thermal problem on the competition between the thermophone and mechanophone contributions are shown in panel (c) of Fig.\ref{fig:histo100ps} and Fig.\ref{fig:histo10ps}, for $\tau_{L}$=100 ps ans 10 ps, respectively.
The striking difference, comparing Fig.\ref{fig:histo1ns}(c), \ref{fig:histo100ps}(c) and \ref{fig:histo10ps}(c), is that the mechanophone contribution dependence on $\tau_{th}/\tau_{L}$ weakens substantially upon reducing the pulse temporal width; not so for the thermophone contribution. Indeed, for the $\tau_{L}$=1 ns case, the mechanophone contribution changes by more than an order of magnitude when increasing $\tau_{TBR}/\tau_{L}$ from 10$^{-2}$ to 10$^2$, whereas it changes by a factor of $\sim$2 for the $\tau_{L}$=10 ps case. The physical reason stands in the thermal problem. The mechanophone effect is triggered by the GNC thermal expansion. The shorter the laser pulse, the smaller is $\delta T_{gnc}$, implying that the peak GNC temperature becomes rather uniform regardless of  $\tau_{TBR}/\tau_{L}$, and the GNC thermal expansion accordingly.\\
\indent When exciting with a 100 ps laser pulse, the mechanophone effect contribution to the radiated acoustic energy in water raises to 23$\%$ for the Au GNC/water interface, and dominates the graphene-functionalized GNC/water interface, see Fig.\ref{fig:histo100ps}(c). Further reducing the pulse duration to 10 ps, the mechanophone effect becomes the prevailing mechanism also for the Au GNC/water interface, see Fig.\ref{fig:histo10ps}(c).\footnote{Note however that, even for the 10 ps light pulse case, the mechanophone effect for $\tau_{TBR}/\tau_L$=10$^{2}$ is lower than the thermophone effect for $\tau_{TBR}/\tau_L$=10$^{-2}$. At room temperature, water thermal expansion coefficient is $\sim$ 5 times higher than the gold's one. Water's expansion is then more efficient than gold's, leading to the maximum of the thermophone effect to exceed that of the mechanophone one. We tested that, using the same thermal expansion coefficients for both the GNC and water results in a maximum of the mechanophone effect (occurring at $\tau_{TBR}/\tau_L$=10$^{-2}$) to be slightly higher than the maximum thermophone effect (occurring at $\tau_{Th}/\tau_L$=10$^{2}$).}\\
\indent In conclusion, we showed, for the case of a water-immersed Au nanocylinder, that the TBR and the laser pulse duration are two valuable control knobs, allowing to switch the acoustic wave launching mechanism in water from the thermophone to the mechanophone. The Au/water and graphene-functionalized Au/water interfaces were discussed as realistic show-case scenarios. Importantly, when the mechanophone is the dominant launching effect, the surrounding water temperature increase is minimized.
These findings thus bare particular importance in situations requiring high frequency acoustic wave generation in water (i.e. short $\tau_L$) while avoiding heating effects of the latter, as is the case for in-vivo bioimaging and theranostics applications at the nanoscale. Noteworthy, these findings may be expanded to include other nanosystems \cite{bertolotti2020note,gandolfi2022ultrafast}.
\section*{Supplementary Material}
See the supplementary material for the details on the simulations design.
\\
\\

\indent This work was partially supported by the LABEX iMUST (ANR-10-LABX-0064) of Université de Lyon within the program "Investissements d'Avenir" (ANR-11-IDEX-0007), by ANR through project 2D-PRESTO (ANR-19-CE09-0027) and through project ULTRASINGLE (ANR-20-CE30-0016).\\
F.B. acknowledges financial support from CNRS through Délégation CNRS 2021-2022.\\
\section*{AUTHOR DECLARATIONS}
\noindent \textbf{Conflict of Interest}\\
\indent The authors have no conflicts to disclose.
\section*{Data Availability Statement}
The data that support the findings of this study are available from the corresponding author upon reasonable request. 
\section*{REFERENCES}
\bibliography{bibliography}

\providecommand{\noopsort}[1]{}\providecommand{\singleletter}[1]{#1}%
\begin{thebibliography}{62}%
\makeatletter
\providecommand \@ifxundefined [1]{%
 \@ifx{#1\undefined}
}%
\providecommand \@ifnum [1]{%
 \ifnum #1\expandafter \@firstoftwo
 \else \expandafter \@secondoftwo
 \fi
}%
\providecommand \@ifx [1]{%
 \ifx #1\expandafter \@firstoftwo
 \else \expandafter \@secondoftwo
 \fi
}%
\providecommand \natexlab [1]{#1}%
\providecommand \enquote  [1]{``#1''}%
\providecommand \bibnamefont  [1]{#1}%
\providecommand \bibfnamefont [1]{#1}%
\providecommand \citenamefont [1]{#1}%
\providecommand \href@noop [0]{\@secondoftwo}%
\providecommand \href [0]{\begingroup \@sanitize@url \@href}%
\providecommand \@href[1]{\@@startlink{#1}\@@href}%
\providecommand \@@href[1]{\endgroup#1\@@endlink}%
\providecommand \@sanitize@url [0]{\catcode `\\12\catcode `\$12\catcode
  `\&12\catcode `\#12\catcode `\^12\catcode `\_12\catcode `\%12\relax}%
\providecommand \@@startlink[1]{}%
\providecommand \@@endlink[0]{}%
\providecommand \url  [0]{\begingroup\@sanitize@url \@url }%
\providecommand \@url [1]{\endgroup\@href {#1}{\urlprefix }}%
\providecommand \urlprefix  [0]{URL }%
\providecommand \Eprint [0]{\href }%
\providecommand \doibase [0]{http://dx.doi.org/}%
\providecommand \selectlanguage [0]{\@gobble}%
\providecommand \bibinfo  [0]{\@secondoftwo}%
\providecommand \bibfield  [0]{\@secondoftwo}%
\providecommand \translation [1]{[#1]}%
\providecommand \BibitemOpen [0]{}%
\providecommand \bibitemStop [0]{}%
\providecommand \bibitemNoStop [0]{.\EOS\space}%
\providecommand \EOS [0]{\spacefactor3000\relax}%
\providecommand \BibitemShut  [1]{\csname bibitem#1\endcsname}%
\let\auto@bib@innerbib\@empty
\bibitem [{\citenamefont {Emelianov}, \citenamefont {Li},\ and\ \citenamefont
  {O’Donnell}(2009)}]{emelianov2009photoacoustics}%
  \BibitemOpen
  \bibfield  {author} {\bibinfo {author} {\bibfnamefont {S.~Y.}\ \bibnamefont
  {Emelianov}}, \bibinfo {author} {\bibfnamefont {P.-C.}\ \bibnamefont {Li}}, \
  and\ \bibinfo {author} {\bibfnamefont {M.}~\bibnamefont {O’Donnell}},\
  }\bibfield  {title} {\enquote {\bibinfo {title} {Photoacoustics for molecular
  imaging and therapy},}\ }\href@noop {} {\bibfield  {journal} {\bibinfo
  {journal} {Physics Today}\ }\textbf {\bibinfo {volume} {62}},\ \bibinfo
  {pages} {34} (\bibinfo {year} {2009})}\BibitemShut {NoStop}%
\bibitem [{\citenamefont {Li}\ and\ \citenamefont
  {Wang}(2009)}]{li2009photoacoustic}%
  \BibitemOpen
  \bibfield  {author} {\bibinfo {author} {\bibfnamefont {C.}~\bibnamefont
  {Li}}\ and\ \bibinfo {author} {\bibfnamefont {L.~V.}\ \bibnamefont {Wang}},\
  }\bibfield  {title} {\enquote {\bibinfo {title} {Photoacoustic tomography and
  sensing in biomedicine},}\ }\href@noop {} {\bibfield  {journal} {\bibinfo
  {journal} {Physics in Medicine \& Biology}\ }\textbf {\bibinfo {volume}
  {54}},\ \bibinfo {pages} {R59} (\bibinfo {year} {2009})}\BibitemShut
  {NoStop}%
\bibitem [{\citenamefont {Xu}\ and\ \citenamefont
  {Wang}(2006)}]{xu2006photoacoustic}%
  \BibitemOpen
  \bibfield  {author} {\bibinfo {author} {\bibfnamefont {M.}~\bibnamefont
  {Xu}}\ and\ \bibinfo {author} {\bibfnamefont {L.~V.}\ \bibnamefont {Wang}},\
  }\bibfield  {title} {\enquote {\bibinfo {title} {Photoacoustic imaging in
  biomedicine},}\ }\href@noop {} {\bibfield  {journal} {\bibinfo  {journal}
  {Review of Scientific Instruments}\ }\textbf {\bibinfo {volume} {77}},\
  \bibinfo {pages} {041101} (\bibinfo {year} {2006})}\BibitemShut {NoStop}%
\bibitem [{\citenamefont {Wang}(2017)}]{wang2017photoacoustic}%
  \BibitemOpen
  \bibfield  {author} {\bibinfo {author} {\bibfnamefont {L.~V.}\ \bibnamefont
  {Wang}},\ }\href@noop {} {\emph {\bibinfo {title} {Photoacoustic imaging and
  spectroscopy}}}\ (\bibinfo  {publisher} {CRC press},\ \bibinfo {year}
  {2017})\BibitemShut {NoStop}%
\bibitem [{\citenamefont {Li}\ and\ \citenamefont {Chen}(2015)}]{li2015gold}%
  \BibitemOpen
  \bibfield  {author} {\bibinfo {author} {\bibfnamefont {W.}~\bibnamefont
  {Li}}\ and\ \bibinfo {author} {\bibfnamefont {X.}~\bibnamefont {Chen}},\
  }\bibfield  {title} {\enquote {\bibinfo {title} {Gold nanoparticles for
  photoacoustic imaging},}\ }\href@noop {} {\bibfield  {journal} {\bibinfo
  {journal} {Nanomedicine}\ }\textbf {\bibinfo {volume} {10}},\ \bibinfo
  {pages} {299--320} (\bibinfo {year} {2015})}\BibitemShut {NoStop}%
\bibitem [{\citenamefont {Yang}\ \emph {et~al.}(2009)\citenamefont {Yang},
  \citenamefont {Stein}, \citenamefont {Ashkenazi},\ and\ \citenamefont
  {Wang}}]{yang2009nanoparticles}%
  \BibitemOpen
  \bibfield  {author} {\bibinfo {author} {\bibfnamefont {X.}~\bibnamefont
  {Yang}}, \bibinfo {author} {\bibfnamefont {E.~W.}\ \bibnamefont {Stein}},
  \bibinfo {author} {\bibfnamefont {S.}~\bibnamefont {Ashkenazi}}, \ and\
  \bibinfo {author} {\bibfnamefont {L.~V.}\ \bibnamefont {Wang}},\ }\bibfield
  {title} {\enquote {\bibinfo {title} {Nanoparticles for photoacoustic
  imaging},}\ }\href@noop {} {\bibfield  {journal} {\bibinfo  {journal} {Wiley
  interdisciplinary reviews: Nanomedicine and Nanobiotechnology}\ }\textbf
  {\bibinfo {volume} {1}},\ \bibinfo {pages} {360--368} (\bibinfo {year}
  {2009})}\BibitemShut {NoStop}%
\bibitem [{\citenamefont {Mantri}\ and\ \citenamefont
  {Jokerst}(2020)}]{mantri2020engineering}%
  \BibitemOpen
  \bibfield  {author} {\bibinfo {author} {\bibfnamefont {Y.}~\bibnamefont
  {Mantri}}\ and\ \bibinfo {author} {\bibfnamefont {J.~V.}\ \bibnamefont
  {Jokerst}},\ }\bibfield  {title} {\enquote {\bibinfo {title} {Engineering
  plasmonic nanoparticles for enhanced photoacoustic imaging},}\ }\href@noop {}
  {\bibfield  {journal} {\bibinfo  {journal} {ACS Nano}\ }\textbf {\bibinfo
  {volume} {14}},\ \bibinfo {pages} {9408--9422} (\bibinfo {year}
  {2020})}\BibitemShut {NoStop}%
\bibitem [{\citenamefont {El-Brolossy}\ \emph {et~al.}(2008)\citenamefont
  {El-Brolossy}, \citenamefont {Abdallah}, \citenamefont {Mohamed},
  \citenamefont {Abdallah}, \citenamefont {Easawi}, \citenamefont {Negm},\ and\
  \citenamefont {Talaat}}]{el2008shape}%
  \BibitemOpen
  \bibfield  {author} {\bibinfo {author} {\bibfnamefont {T.}~\bibnamefont
  {El-Brolossy}}, \bibinfo {author} {\bibfnamefont {T.}~\bibnamefont
  {Abdallah}}, \bibinfo {author} {\bibfnamefont {M.~B.}\ \bibnamefont
  {Mohamed}}, \bibinfo {author} {\bibfnamefont {S.}~\bibnamefont {Abdallah}},
  \bibinfo {author} {\bibfnamefont {K.}~\bibnamefont {Easawi}}, \bibinfo
  {author} {\bibfnamefont {S.}~\bibnamefont {Negm}}, \ and\ \bibinfo {author}
  {\bibfnamefont {H.}~\bibnamefont {Talaat}},\ }\bibfield  {title} {\enquote
  {\bibinfo {title} {Shape and size dependence of the surface plasmon resonance
  of gold nanoparticles studied by photoacoustic technique},}\ }\href@noop {}
  {\bibfield  {journal} {\bibinfo  {journal} {The European Physical Journal
  Special Topics}\ }\textbf {\bibinfo {volume} {153}},\ \bibinfo {pages}
  {361--364} (\bibinfo {year} {2008})}\BibitemShut {NoStop}%
\bibitem [{\citenamefont {Garcia-Alvarez}\ \emph {et~al.}(2020)\citenamefont
  {Garcia-Alvarez}, \citenamefont {Chen}, \citenamefont {Nedilko},
  \citenamefont {S{\'a}nchez-Iglesias}, \citenamefont {Rix}, \citenamefont
  {Lederle}, \citenamefont {Pathak}, \citenamefont {Lammers}, \citenamefont
  {Von~Plessen}, \citenamefont {Kostarelos}, \citenamefont {Liz-Marzán},
  \citenamefont {Kuehne},\ and\ \citenamefont
  {Chigrin}}]{garcia2020optimizing}%
  \BibitemOpen
  \bibfield  {author} {\bibinfo {author} {\bibfnamefont {R.}~\bibnamefont
  {Garcia-Alvarez}}, \bibinfo {author} {\bibfnamefont {L.}~\bibnamefont
  {Chen}}, \bibinfo {author} {\bibfnamefont {A.}~\bibnamefont {Nedilko}},
  \bibinfo {author} {\bibfnamefont {A.}~\bibnamefont {S{\'a}nchez-Iglesias}},
  \bibinfo {author} {\bibfnamefont {A.}~\bibnamefont {Rix}}, \bibinfo {author}
  {\bibfnamefont {W.}~\bibnamefont {Lederle}}, \bibinfo {author} {\bibfnamefont
  {V.}~\bibnamefont {Pathak}}, \bibinfo {author} {\bibfnamefont
  {T.}~\bibnamefont {Lammers}}, \bibinfo {author} {\bibfnamefont
  {G.}~\bibnamefont {Von~Plessen}}, \bibinfo {author} {\bibfnamefont
  {K.}~\bibnamefont {Kostarelos}}, \bibinfo {author} {\bibfnamefont {L.~M.}\
  \bibnamefont {Liz-Marzán}}, \bibinfo {author} {\bibfnamefont {A.~J.}\
  \bibnamefont {Kuehne}}, \ and\ \bibinfo {author} {\bibfnamefont {D.~N.}\
  \bibnamefont {Chigrin}},\ }\bibfield  {title} {\enquote {\bibinfo {title}
  {Optimizing the geometry of photoacoustically active gold nanoparticles for
  biomedical imaging},}\ }\href@noop {} {\bibfield  {journal} {\bibinfo
  {journal} {ACS Photonics}\ }\textbf {\bibinfo {volume} {7}},\ \bibinfo
  {pages} {646--652} (\bibinfo {year} {2020})}\BibitemShut {NoStop}%
\bibitem [{\citenamefont {Luke}, \citenamefont {Yeager},\ and\ \citenamefont
  {Emelianov}(2012)}]{luke2012biomedical}%
  \BibitemOpen
  \bibfield  {author} {\bibinfo {author} {\bibfnamefont {G.~P.}\ \bibnamefont
  {Luke}}, \bibinfo {author} {\bibfnamefont {D.}~\bibnamefont {Yeager}}, \ and\
  \bibinfo {author} {\bibfnamefont {S.~Y.}\ \bibnamefont {Emelianov}},\
  }\bibfield  {title} {\enquote {\bibinfo {title} {Biomedical applications of
  photoacoustic imaging with exogenous contrast agents},}\ }\href@noop {}
  {\bibfield  {journal} {\bibinfo  {journal} {Annals of Biomedical
  Engineering}\ }\textbf {\bibinfo {volume} {40}},\ \bibinfo {pages} {422--437}
  (\bibinfo {year} {2012})}\BibitemShut {NoStop}%
\bibitem [{\citenamefont {Pan}\ \emph {et~al.}(2011)\citenamefont {Pan},
  \citenamefont {Pramanik}, \citenamefont {Senpan}, \citenamefont {Allen},
  \citenamefont {Zhang}, \citenamefont {Wickline}, \citenamefont {Wang},\ and\
  \citenamefont {Lanza}}]{pan2011molecular}%
  \BibitemOpen
  \bibfield  {author} {\bibinfo {author} {\bibfnamefont {D.}~\bibnamefont
  {Pan}}, \bibinfo {author} {\bibfnamefont {M.}~\bibnamefont {Pramanik}},
  \bibinfo {author} {\bibfnamefont {A.}~\bibnamefont {Senpan}}, \bibinfo
  {author} {\bibfnamefont {J.~S.}\ \bibnamefont {Allen}}, \bibinfo {author}
  {\bibfnamefont {H.}~\bibnamefont {Zhang}}, \bibinfo {author} {\bibfnamefont
  {S.~A.}\ \bibnamefont {Wickline}}, \bibinfo {author} {\bibfnamefont {L.~V.}\
  \bibnamefont {Wang}}, \ and\ \bibinfo {author} {\bibfnamefont {G.~M.}\
  \bibnamefont {Lanza}},\ }\bibfield  {title} {\enquote {\bibinfo {title}
  {Molecular photoacoustic imaging of angiogenesis with integrin-targeted gold
  nanobeacons},}\ }\href@noop {} {\bibfield  {journal} {\bibinfo  {journal}
  {The FASEB Journal}\ }\textbf {\bibinfo {volume} {25}},\ \bibinfo {pages}
  {875--882} (\bibinfo {year} {2011})}\BibitemShut {NoStop}%
\bibitem [{\citenamefont {Shukla}\ \emph {et~al.}(2005)\citenamefont {Shukla},
  \citenamefont {Bansal}, \citenamefont {Chaudhary}, \citenamefont {Basu},
  \citenamefont {Bhonde},\ and\ \citenamefont
  {Sastry}}]{shukla2005biocompatibility}%
  \BibitemOpen
  \bibfield  {author} {\bibinfo {author} {\bibfnamefont {R.}~\bibnamefont
  {Shukla}}, \bibinfo {author} {\bibfnamefont {V.}~\bibnamefont {Bansal}},
  \bibinfo {author} {\bibfnamefont {M.}~\bibnamefont {Chaudhary}}, \bibinfo
  {author} {\bibfnamefont {A.}~\bibnamefont {Basu}}, \bibinfo {author}
  {\bibfnamefont {R.~R.}\ \bibnamefont {Bhonde}}, \ and\ \bibinfo {author}
  {\bibfnamefont {M.}~\bibnamefont {Sastry}},\ }\bibfield  {title} {\enquote
  {\bibinfo {title} {Biocompatibility of gold nanoparticles and their
  endocytotic fate inside the cellular compartment: a microscopic overview},}\
  }\href@noop {} {\bibfield  {journal} {\bibinfo  {journal} {Langmuir}\
  }\textbf {\bibinfo {volume} {21}},\ \bibinfo {pages} {10644--10654} (\bibinfo
  {year} {2005})}\BibitemShut {NoStop}%
\bibitem [{\citenamefont {Craciun}\ \emph {et~al.}(2017)\citenamefont
  {Craciun}, \citenamefont {Focsan}, \citenamefont {Magyari}, \citenamefont
  {Vulpoi},\ and\ \citenamefont {Pap}}]{craciun2017surface}%
  \BibitemOpen
  \bibfield  {author} {\bibinfo {author} {\bibfnamefont {A.~M.}\ \bibnamefont
  {Craciun}}, \bibinfo {author} {\bibfnamefont {M.}~\bibnamefont {Focsan}},
  \bibinfo {author} {\bibfnamefont {K.}~\bibnamefont {Magyari}}, \bibinfo
  {author} {\bibfnamefont {A.}~\bibnamefont {Vulpoi}}, \ and\ \bibinfo {author}
  {\bibfnamefont {Z.}~\bibnamefont {Pap}},\ }\bibfield  {title} {\enquote
  {\bibinfo {title} {Surface plasmon resonance or biocompatibility—key
  properties for determining the applicability of noble metal nanoparticles},}\
  }\href@noop {} {\bibfield  {journal} {\bibinfo  {journal} {Materials}\
  }\textbf {\bibinfo {volume} {10}},\ \bibinfo {pages} {836} (\bibinfo {year}
  {2017})}\BibitemShut {NoStop}%
\bibitem [{\citenamefont {Verawaty}\ and\ \citenamefont
  {Pramanik}(2016)}]{pramanik2016simulating}%
  \BibitemOpen
  \bibfield  {author} {\bibinfo {author} {\bibnamefont {Verawaty}}\ and\
  \bibinfo {author} {\bibfnamefont {M.}~\bibnamefont {Pramanik}},\ }\bibfield
  {title} {\enquote {\bibinfo {title} {Simulating photoacoustic waves from
  individual nanoparticle of various shapes using k-wave},}\ }\href@noop {}
  {\bibfield  {journal} {\bibinfo  {journal} {Biomedical Physics \& Engineering
  Express}\ }\textbf {\bibinfo {volume} {2}},\ \bibinfo {pages} {035013}
  (\bibinfo {year} {2016})}\BibitemShut {NoStop}%
\bibitem [{\citenamefont {Guiraud}\ \emph {et~al.}(2021)\citenamefont
  {Guiraud}, \citenamefont {Giordano}, \citenamefont {Bou~Matar}, \citenamefont
  {Pernod},\ and\ \citenamefont {Lardat}}]{guiraud2021thermoacoustic}%
  \BibitemOpen
  \bibfield  {author} {\bibinfo {author} {\bibfnamefont {P.}~\bibnamefont
  {Guiraud}}, \bibinfo {author} {\bibfnamefont {S.}~\bibnamefont {Giordano}},
  \bibinfo {author} {\bibfnamefont {O.}~\bibnamefont {Bou~Matar}}, \bibinfo
  {author} {\bibfnamefont {P.}~\bibnamefont {Pernod}}, \ and\ \bibinfo {author}
  {\bibfnamefont {R.}~\bibnamefont {Lardat}},\ }\bibfield  {title} {\enquote
  {\bibinfo {title} {Thermoacoustic wave generation in multilayered
  thermophones with cylindrical and spherical geometries},}\ }\href@noop {}
  {\bibfield  {journal} {\bibinfo  {journal} {Journal of Applied Physics}\
  }\textbf {\bibinfo {volume} {129}},\ \bibinfo {pages} {115103} (\bibinfo
  {year} {2021})}\BibitemShut {NoStop}%
\bibitem [{\citenamefont {Ronchi}\ \emph {et~al.}(2021)\citenamefont {Ronchi},
  \citenamefont {Sterzi}, \citenamefont {Gandolfi}, \citenamefont {Belarouci},
  \citenamefont {Giannetti}, \citenamefont {Del~Fatti}, \citenamefont {Banfi},\
  and\ \citenamefont {Ferrini}}]{ronchi2021discrimination}%
  \BibitemOpen
  \bibfield  {author} {\bibinfo {author} {\bibfnamefont {A.}~\bibnamefont
  {Ronchi}}, \bibinfo {author} {\bibfnamefont {A.}~\bibnamefont {Sterzi}},
  \bibinfo {author} {\bibfnamefont {M.}~\bibnamefont {Gandolfi}}, \bibinfo
  {author} {\bibfnamefont {A.}~\bibnamefont {Belarouci}}, \bibinfo {author}
  {\bibfnamefont {C.}~\bibnamefont {Giannetti}}, \bibinfo {author}
  {\bibfnamefont {N.}~\bibnamefont {Del~Fatti}}, \bibinfo {author}
  {\bibfnamefont {F.}~\bibnamefont {Banfi}}, \ and\ \bibinfo {author}
  {\bibfnamefont {G.}~\bibnamefont {Ferrini}},\ }\bibfield  {title} {\enquote
  {\bibinfo {title} {Discrimination of nano-objects via cluster analysis
  techniques applied to time-resolved thermo-acoustic microscopy},}\
  }\href@noop {} {\bibfield  {journal} {\bibinfo  {journal} {Ultrasonics}\
  }\textbf {\bibinfo {volume} {114}},\ \bibinfo {pages} {106403} (\bibinfo
  {year} {2021})}\BibitemShut {NoStop}%
\bibitem [{\citenamefont {Lee}\ \emph {et~al.}(2018)\citenamefont {Lee},
  \citenamefont {Baac}, \citenamefont {Li},\ and\ \citenamefont
  {Guo}}]{lee2018efficient}%
  \BibitemOpen
  \bibfield  {author} {\bibinfo {author} {\bibfnamefont {T.}~\bibnamefont
  {Lee}}, \bibinfo {author} {\bibfnamefont {H.~W.}\ \bibnamefont {Baac}},
  \bibinfo {author} {\bibfnamefont {Q.}~\bibnamefont {Li}}, \ and\ \bibinfo
  {author} {\bibfnamefont {L.~J.}\ \bibnamefont {Guo}},\ }\bibfield  {title}
  {\enquote {\bibinfo {title} {Efficient photoacoustic conversion in optical
  nanomaterials and composites},}\ }\href@noop {} {\bibfield  {journal}
  {\bibinfo  {journal} {Advanced Optical Materials}\ }\textbf {\bibinfo
  {volume} {6}},\ \bibinfo {pages} {1800491} (\bibinfo {year}
  {2018})}\BibitemShut {NoStop}%
\bibitem [{\citenamefont {Gandolfi}, \citenamefont {Banfi},\ and\ \citenamefont
  {Glorieux}(2020)}]{gandolfi2020optical}%
  \BibitemOpen
  \bibfield  {author} {\bibinfo {author} {\bibfnamefont {M.}~\bibnamefont
  {Gandolfi}}, \bibinfo {author} {\bibfnamefont {F.}~\bibnamefont {Banfi}}, \
  and\ \bibinfo {author} {\bibfnamefont {C.}~\bibnamefont {Glorieux}},\
  }\bibfield  {title} {\enquote {\bibinfo {title} {Optical wavelength
  dependence of photoacoustic signal of gold nanofluid},}\ }\href@noop {}
  {\bibfield  {journal} {\bibinfo  {journal} {Photoacoustics}\ }\textbf
  {\bibinfo {volume} {20}},\ \bibinfo {pages} {100199} (\bibinfo {year}
  {2020})}\BibitemShut {NoStop}%
\bibitem [{\citenamefont {Prost}, \citenamefont {Poisson},\ and\ \citenamefont
  {Bossy}(2015)}]{prost2015photoacoustic}%
  \BibitemOpen
  \bibfield  {author} {\bibinfo {author} {\bibfnamefont {A.}~\bibnamefont
  {Prost}}, \bibinfo {author} {\bibfnamefont {F.}~\bibnamefont {Poisson}}, \
  and\ \bibinfo {author} {\bibfnamefont {E.}~\bibnamefont {Bossy}},\ }\bibfield
   {title} {\enquote {\bibinfo {title} {Photoacoustic generation by a gold
  nanosphere: From linear to nonlinear thermoelastics in the long-pulse
  illumination regime},}\ }\href@noop {} {\bibfield  {journal} {\bibinfo
  {journal} {Physical Review B}\ }\textbf {\bibinfo {volume} {92}},\ \bibinfo
  {pages} {115450} (\bibinfo {year} {2015})}\BibitemShut {NoStop}%
\bibitem [{\citenamefont {Diego}\ \emph {et~al.}(2022)\citenamefont {Diego},
  \citenamefont {Gandolfi}, \citenamefont {Casto}, \citenamefont
  {Maria~Bellussi}, \citenamefont {Vialla}, \citenamefont {Crut}, \citenamefont
  {Roddaro}, \citenamefont {Fasano}, \citenamefont {Vall\'{e}e}, \citenamefont
  {Del~Fatti}, \citenamefont {Maioli},\ and\ \citenamefont
  {Banfi}}]{diego2022Ultrafast}%
  \BibitemOpen
  \bibfield  {author} {\bibinfo {author} {\bibfnamefont {M.}~\bibnamefont
  {Diego}}, \bibinfo {author} {\bibfnamefont {M.}~\bibnamefont {Gandolfi}},
  \bibinfo {author} {\bibfnamefont {A.}~\bibnamefont {Casto}}, \bibinfo
  {author} {\bibfnamefont {F.}~\bibnamefont {Maria~Bellussi}}, \bibinfo
  {author} {\bibfnamefont {F.}~\bibnamefont {Vialla}}, \bibinfo {author}
  {\bibfnamefont {A.}~\bibnamefont {Crut}}, \bibinfo {author} {\bibfnamefont
  {S.}~\bibnamefont {Roddaro}}, \bibinfo {author} {\bibfnamefont
  {M.}~\bibnamefont {Fasano}}, \bibinfo {author} {\bibfnamefont
  {F.}~\bibnamefont {Vall\'{e}e}}, \bibinfo {author} {\bibfnamefont
  {N.}~\bibnamefont {Del~Fatti}}, \bibinfo {author} {\bibfnamefont
  {P.}~\bibnamefont {Maioli}}, \ and\ \bibinfo {author} {\bibfnamefont
  {F.}~\bibnamefont {Banfi}},\ }\bibfield  {title} {\enquote {\bibinfo {title}
  {Ultrafast nano generation of acoustic waves in water via a single carbon
  nanotube},}\ }\href@noop {} {\bibfield  {journal} {\bibinfo  {journal}
  {Photoacoustics}\ ,\ \bibinfo {pages} {100407}} (\bibinfo {year}
  {2022})}\BibitemShut {NoStop}%
\bibitem [{\citenamefont {Chen}\ \emph {et~al.}(2011)\citenamefont {Chen},
  \citenamefont {Frey}, \citenamefont {Kim}, \citenamefont {Kruizinga},
  \citenamefont {Homan},\ and\ \citenamefont {Emelianov}}]{chen2011silica}%
  \BibitemOpen
  \bibfield  {author} {\bibinfo {author} {\bibfnamefont {Y.-S.}\ \bibnamefont
  {Chen}}, \bibinfo {author} {\bibfnamefont {W.}~\bibnamefont {Frey}}, \bibinfo
  {author} {\bibfnamefont {S.}~\bibnamefont {Kim}}, \bibinfo {author}
  {\bibfnamefont {P.}~\bibnamefont {Kruizinga}}, \bibinfo {author}
  {\bibfnamefont {K.}~\bibnamefont {Homan}}, \ and\ \bibinfo {author}
  {\bibfnamefont {S.}~\bibnamefont {Emelianov}},\ }\bibfield  {title} {\enquote
  {\bibinfo {title} {Silica-coated gold nanorods as photoacoustic signal
  nanoamplifiers},}\ }\href@noop {} {\bibfield  {journal} {\bibinfo  {journal}
  {Nano Letters}\ }\textbf {\bibinfo {volume} {11}},\ \bibinfo {pages}
  {348--354} (\bibinfo {year} {2011})}\BibitemShut {NoStop}%
\bibitem [{\citenamefont {Repenko}\ \emph {et~al.}(2018)\citenamefont
  {Repenko}, \citenamefont {Rix}, \citenamefont {Nedilko}, \citenamefont
  {Rose}, \citenamefont {Hermann}, \citenamefont {Vinokur}, \citenamefont
  {Moli}, \citenamefont {Cao-Mil{\`a}n}, \citenamefont {Mayer}, \citenamefont
  {von Plessen}, \citenamefont {Fery}, \citenamefont {De~Laporte},
  \citenamefont {Lederle}, \citenamefont {Chigrin},\ and\ \citenamefont
  {Kuehne}}]{repenko2018strong}%
  \BibitemOpen
  \bibfield  {author} {\bibinfo {author} {\bibfnamefont {T.}~\bibnamefont
  {Repenko}}, \bibinfo {author} {\bibfnamefont {A.}~\bibnamefont {Rix}},
  \bibinfo {author} {\bibfnamefont {A.}~\bibnamefont {Nedilko}}, \bibinfo
  {author} {\bibfnamefont {J.}~\bibnamefont {Rose}}, \bibinfo {author}
  {\bibfnamefont {A.}~\bibnamefont {Hermann}}, \bibinfo {author} {\bibfnamefont
  {R.}~\bibnamefont {Vinokur}}, \bibinfo {author} {\bibfnamefont
  {S.}~\bibnamefont {Moli}}, \bibinfo {author} {\bibfnamefont {R.}~\bibnamefont
  {Cao-Mil{\`a}n}}, \bibinfo {author} {\bibfnamefont {M.}~\bibnamefont
  {Mayer}}, \bibinfo {author} {\bibfnamefont {G.}~\bibnamefont {von Plessen}},
  \bibinfo {author} {\bibfnamefont {A.}~\bibnamefont {Fery}}, \bibinfo {author}
  {\bibfnamefont {L.}~\bibnamefont {De~Laporte}}, \bibinfo {author}
  {\bibfnamefont {W.}~\bibnamefont {Lederle}}, \bibinfo {author} {\bibfnamefont
  {D.~N.}\ \bibnamefont {Chigrin}}, \ and\ \bibinfo {author} {\bibfnamefont
  {A.~J.~C.}\ \bibnamefont {Kuehne}},\ }\bibfield  {title} {\enquote {\bibinfo
  {title} {Strong photoacoustic signal enhancement by coating gold
  nanoparticles with melanin for biomedical imaging},}\ }\href@noop {}
  {\bibfield  {journal} {\bibinfo  {journal} {Advanced Functional Materials}\
  }\textbf {\bibinfo {volume} {28}},\ \bibinfo {pages} {1705607} (\bibinfo
  {year} {2018})}\BibitemShut {NoStop}%
\bibitem [{\citenamefont {Cavigli}\ \emph {et~al.}(2020)\citenamefont
  {Cavigli}, \citenamefont {Milanesi}, \citenamefont {Khlebtsov}, \citenamefont
  {Centi}, \citenamefont {Ratto}, \citenamefont {Khlebtsov},\ and\
  \citenamefont {Pini}}]{cavigli2020impact}%
  \BibitemOpen
  \bibfield  {author} {\bibinfo {author} {\bibfnamefont {L.}~\bibnamefont
  {Cavigli}}, \bibinfo {author} {\bibfnamefont {A.}~\bibnamefont {Milanesi}},
  \bibinfo {author} {\bibfnamefont {B.~N.}\ \bibnamefont {Khlebtsov}}, \bibinfo
  {author} {\bibfnamefont {S.}~\bibnamefont {Centi}}, \bibinfo {author}
  {\bibfnamefont {F.}~\bibnamefont {Ratto}}, \bibinfo {author} {\bibfnamefont
  {N.~G.}\ \bibnamefont {Khlebtsov}}, \ and\ \bibinfo {author} {\bibfnamefont
  {R.}~\bibnamefont {Pini}},\ }\bibfield  {title} {\enquote {\bibinfo {title}
  {Impact of kapitza resistance on the stability and efficiency of
  photoacoustic conversion from gold nanorods},}\ }\href@noop {} {\bibfield
  {journal} {\bibinfo  {journal} {Journal of Colloid and Interface Science}\
  }\textbf {\bibinfo {volume} {578}},\ \bibinfo {pages} {358--365} (\bibinfo
  {year} {2020})}\BibitemShut {NoStop}%
\bibitem [{\citenamefont {Chen}\ \emph {et~al.}(2012)\citenamefont {Chen},
  \citenamefont {Frey}, \citenamefont {Aglyamov},\ and\ \citenamefont
  {Emelianov}}]{chen2012environment}%
  \BibitemOpen
  \bibfield  {author} {\bibinfo {author} {\bibfnamefont {Y.-S.}\ \bibnamefont
  {Chen}}, \bibinfo {author} {\bibfnamefont {W.}~\bibnamefont {Frey}}, \bibinfo
  {author} {\bibfnamefont {S.}~\bibnamefont {Aglyamov}}, \ and\ \bibinfo
  {author} {\bibfnamefont {S.}~\bibnamefont {Emelianov}},\ }\bibfield  {title}
  {\enquote {\bibinfo {title} {Environment-dependent generation of
  photoacoustic waves from plasmonic nanoparticles},}\ }\href@noop {}
  {\bibfield  {journal} {\bibinfo  {journal} {Small}\ }\textbf {\bibinfo
  {volume} {8}},\ \bibinfo {pages} {47--52} (\bibinfo {year}
  {2012})}\BibitemShut {NoStop}%
\bibitem [{\citenamefont {Shahbazi}\ \emph {et~al.}(2019)\citenamefont
  {Shahbazi}, \citenamefont {Frey}, \citenamefont {Chen}, \citenamefont
  {Aglyamov},\ and\ \citenamefont {Emelianov}}]{shahbazi2019photoacoustics}%
  \BibitemOpen
  \bibfield  {author} {\bibinfo {author} {\bibfnamefont {K.}~\bibnamefont
  {Shahbazi}}, \bibinfo {author} {\bibfnamefont {W.}~\bibnamefont {Frey}},
  \bibinfo {author} {\bibfnamefont {Y.-S.}\ \bibnamefont {Chen}}, \bibinfo
  {author} {\bibfnamefont {S.}~\bibnamefont {Aglyamov}}, \ and\ \bibinfo
  {author} {\bibfnamefont {S.}~\bibnamefont {Emelianov}},\ }\bibfield  {title}
  {\enquote {\bibinfo {title} {Photoacoustics of core--shell nanospheres using
  comprehensive modeling and analytical solution approach},}\ }\href@noop {}
  {\bibfield  {journal} {\bibinfo  {journal} {Communications Physics}\ }\textbf
  {\bibinfo {volume} {2}},\ \bibinfo {pages} {1--11} (\bibinfo {year}
  {2019})}\BibitemShut {NoStop}%
\bibitem [{\citenamefont {Guiraud}\ \emph
  {et~al.}(2019{\natexlab{a}})\citenamefont {Guiraud}, \citenamefont
  {Giordano}, \citenamefont {Bou-Matar}, \citenamefont {Pernod},\ and\
  \citenamefont {Lardat}}]{guiraud2019multilayer}%
  \BibitemOpen
  \bibfield  {author} {\bibinfo {author} {\bibfnamefont {P.}~\bibnamefont
  {Guiraud}}, \bibinfo {author} {\bibfnamefont {S.}~\bibnamefont {Giordano}},
  \bibinfo {author} {\bibfnamefont {O.}~\bibnamefont {Bou-Matar}}, \bibinfo
  {author} {\bibfnamefont {P.}~\bibnamefont {Pernod}}, \ and\ \bibinfo {author}
  {\bibfnamefont {R.}~\bibnamefont {Lardat}},\ }\bibfield  {title} {\enquote
  {\bibinfo {title} {Multilayer modeling of thermoacoustic sound generation for
  thermophone analysis and design},}\ }\href@noop {} {\bibfield  {journal}
  {\bibinfo  {journal} {Journal of Sound and Vibration}\ }\textbf {\bibinfo
  {volume} {455}},\ \bibinfo {pages} {275--298} (\bibinfo {year}
  {2019}{\natexlab{a}})}\BibitemShut {NoStop}%
\bibitem [{\citenamefont {Guiraud}\ \emph
  {et~al.}(2019{\natexlab{b}})\citenamefont {Guiraud}, \citenamefont
  {Giordano}, \citenamefont {Bou~Matar}, \citenamefont {Pernod},\ and\
  \citenamefont {Lardat}}]{guiraud2019two}%
  \BibitemOpen
  \bibfield  {author} {\bibinfo {author} {\bibfnamefont {P.}~\bibnamefont
  {Guiraud}}, \bibinfo {author} {\bibfnamefont {S.}~\bibnamefont {Giordano}},
  \bibinfo {author} {\bibfnamefont {O.}~\bibnamefont {Bou~Matar}}, \bibinfo
  {author} {\bibfnamefont {P.}~\bibnamefont {Pernod}}, \ and\ \bibinfo {author}
  {\bibfnamefont {R.}~\bibnamefont {Lardat}},\ }\bibfield  {title} {\enquote
  {\bibinfo {title} {Two temperature model for thermoacoustic sound generation
  in thick porous thermophones},}\ }\href@noop {} {\bibfield  {journal}
  {\bibinfo  {journal} {Journal of Applied Physics}\ }\textbf {\bibinfo
  {volume} {126}},\ \bibinfo {pages} {165111} (\bibinfo {year}
  {2019}{\natexlab{b}})}\BibitemShut {NoStop}%
\bibitem [{\citenamefont {He}\ \emph {et~al.}(2020)\citenamefont {He},
  \citenamefont {Wang}, \citenamefont {Gao}, \citenamefont {Lu},\ and\
  \citenamefont {Song}}]{he2020application}%
  \BibitemOpen
  \bibfield  {author} {\bibinfo {author} {\bibfnamefont {W.}~\bibnamefont
  {He}}, \bibinfo {author} {\bibfnamefont {X.}~\bibnamefont {Wang}}, \bibinfo
  {author} {\bibfnamefont {X.}~\bibnamefont {Gao}}, \bibinfo {author}
  {\bibfnamefont {Z.}~\bibnamefont {Lu}}, \ and\ \bibinfo {author}
  {\bibfnamefont {J.}~\bibnamefont {Song}},\ }\bibfield  {title} {\enquote
  {\bibinfo {title} {Application of gold nanoparticles in photoacoustic
  imaging},}\ }in\ \href@noop {} {\emph {\bibinfo {booktitle} {IOP Conference
  Series: Materials Science and Engineering}}},\ Vol.\ \bibinfo {volume} {729}\
  (\bibinfo {organization} {IOP Publishing},\ \bibinfo {year} {2020})\ p.\
  \bibinfo {pages} {012086}\BibitemShut {NoStop}%
\bibitem [{\citenamefont {Manohar}, \citenamefont {Ungureanu},\ and\
  \citenamefont {Van~Leeuwen}(2011)}]{manohar2011gold}%
  \BibitemOpen
  \bibfield  {author} {\bibinfo {author} {\bibfnamefont {S.}~\bibnamefont
  {Manohar}}, \bibinfo {author} {\bibfnamefont {C.}~\bibnamefont {Ungureanu}},
  \ and\ \bibinfo {author} {\bibfnamefont {T.~G.}\ \bibnamefont
  {Van~Leeuwen}},\ }\bibfield  {title} {\enquote {\bibinfo {title} {Gold
  nanorods as molecular contrast agents in photoacoustic imaging: the promises
  and the caveats},}\ }\href@noop {} {\bibfield  {journal} {\bibinfo  {journal}
  {Contrast Media \& Molecular Imaging}\ }\textbf {\bibinfo {volume} {6}},\
  \bibinfo {pages} {389--400} (\bibinfo {year} {2011})}\BibitemShut {NoStop}%
\bibitem [{\citenamefont {Cavigli}\ \emph {et~al.}(2014)\citenamefont
  {Cavigli}, \citenamefont {de~Angelis}, \citenamefont {Ratto}, \citenamefont
  {Matteini}, \citenamefont {Rossi}, \citenamefont {Centi}, \citenamefont
  {Fusi},\ and\ \citenamefont {Pini}}]{cavigli2014size}%
  \BibitemOpen
  \bibfield  {author} {\bibinfo {author} {\bibfnamefont {L.}~\bibnamefont
  {Cavigli}}, \bibinfo {author} {\bibfnamefont {M.}~\bibnamefont {de~Angelis}},
  \bibinfo {author} {\bibfnamefont {F.}~\bibnamefont {Ratto}}, \bibinfo
  {author} {\bibfnamefont {P.}~\bibnamefont {Matteini}}, \bibinfo {author}
  {\bibfnamefont {F.}~\bibnamefont {Rossi}}, \bibinfo {author} {\bibfnamefont
  {S.}~\bibnamefont {Centi}}, \bibinfo {author} {\bibfnamefont
  {F.}~\bibnamefont {Fusi}}, \ and\ \bibinfo {author} {\bibfnamefont
  {R.}~\bibnamefont {Pini}},\ }\bibfield  {title} {\enquote {\bibinfo {title}
  {Size affects the stability of the photoacoustic conversion of gold
  nanorods},}\ }\href@noop {} {\bibfield  {journal} {\bibinfo  {journal} {The
  Journal of Physical Chemistry C}\ }\textbf {\bibinfo {volume} {118}},\
  \bibinfo {pages} {16140--16146} (\bibinfo {year} {2014})}\BibitemShut
  {NoStop}%
\bibitem [{\citenamefont {Cavigli}\ \emph {et~al.}(2017)\citenamefont
  {Cavigli}, \citenamefont {Cini}, \citenamefont {Centi}, \citenamefont
  {Borri}, \citenamefont {Lai}, \citenamefont {Ratto}, \citenamefont
  {de~Angelis},\ and\ \citenamefont {Pini}}]{cavigli2017photo}%
  \BibitemOpen
  \bibfield  {author} {\bibinfo {author} {\bibfnamefont {L.}~\bibnamefont
  {Cavigli}}, \bibinfo {author} {\bibfnamefont {A.}~\bibnamefont {Cini}},
  \bibinfo {author} {\bibfnamefont {S.}~\bibnamefont {Centi}}, \bibinfo
  {author} {\bibfnamefont {C.}~\bibnamefont {Borri}}, \bibinfo {author}
  {\bibfnamefont {S.}~\bibnamefont {Lai}}, \bibinfo {author} {\bibfnamefont
  {F.}~\bibnamefont {Ratto}}, \bibinfo {author} {\bibfnamefont
  {M.}~\bibnamefont {de~Angelis}}, \ and\ \bibinfo {author} {\bibfnamefont
  {R.}~\bibnamefont {Pini}},\ }\bibfield  {title} {\enquote {\bibinfo {title}
  {Photostability of gold nanorods upon endosomal confinement in cultured
  cells},}\ }\href@noop {} {\bibfield  {journal} {\bibinfo  {journal} {The
  Journal of Physical Chemistry C}\ }\textbf {\bibinfo {volume} {121}},\
  \bibinfo {pages} {6393--6400} (\bibinfo {year} {2017})}\BibitemShut {NoStop}%
\bibitem [{\citenamefont {Chen}\ \emph {et~al.}(2019)\citenamefont {Chen},
  \citenamefont {Zhao}, \citenamefont {Yoon}, \citenamefont {Gambhir},\ and\
  \citenamefont {Emelianov}}]{chen2019miniature}%
  \BibitemOpen
  \bibfield  {author} {\bibinfo {author} {\bibfnamefont {Y.-S.}\ \bibnamefont
  {Chen}}, \bibinfo {author} {\bibfnamefont {Y.}~\bibnamefont {Zhao}}, \bibinfo
  {author} {\bibfnamefont {S.~J.}\ \bibnamefont {Yoon}}, \bibinfo {author}
  {\bibfnamefont {S.~S.}\ \bibnamefont {Gambhir}}, \ and\ \bibinfo {author}
  {\bibfnamefont {S.}~\bibnamefont {Emelianov}},\ }\bibfield  {title} {\enquote
  {\bibinfo {title} {Miniature gold nanorods for photoacoustic molecular
  imaging in the second near-infrared optical window},}\ }\href@noop {}
  {\bibfield  {journal} {\bibinfo  {journal} {Nature Nanotechnology}\ }\textbf
  {\bibinfo {volume} {14}},\ \bibinfo {pages} {465--472} (\bibinfo {year}
  {2019})}\BibitemShut {NoStop}%
\bibitem [{\citenamefont {Dhada}\ \emph {et~al.}(2020)\citenamefont {Dhada},
  \citenamefont {Hernandez}, \citenamefont {Huang},\ and\ \citenamefont
  {Suggs}}]{dhada2020gold}%
  \BibitemOpen
  \bibfield  {author} {\bibinfo {author} {\bibfnamefont {K.~S.}\ \bibnamefont
  {Dhada}}, \bibinfo {author} {\bibfnamefont {D.~S.}\ \bibnamefont
  {Hernandez}}, \bibinfo {author} {\bibfnamefont {W.}~\bibnamefont {Huang}}, \
  and\ \bibinfo {author} {\bibfnamefont {L.~J.}\ \bibnamefont {Suggs}},\
  }\bibfield  {title} {\enquote {\bibinfo {title} {Gold nanorods as
  photoacoustic nanoprobes to detect proinflammatory macrophages and
  inflammation},}\ }\href@noop {} {\bibfield  {journal} {\bibinfo  {journal}
  {ACS Applied Nano Materials}\ }\textbf {\bibinfo {volume} {3}},\ \bibinfo
  {pages} {7774--7780} (\bibinfo {year} {2020})}\BibitemShut {NoStop}%
\bibitem [{\citenamefont {Knights}\ \emph {et~al.}(2019)\citenamefont
  {Knights}, \citenamefont {Ye}, \citenamefont {Ingram}, \citenamefont
  {Freear},\ and\ \citenamefont {McLaughlan}}]{knights2019optimising}%
  \BibitemOpen
  \bibfield  {author} {\bibinfo {author} {\bibfnamefont {O.~B.}\ \bibnamefont
  {Knights}}, \bibinfo {author} {\bibfnamefont {S.}~\bibnamefont {Ye}},
  \bibinfo {author} {\bibfnamefont {N.}~\bibnamefont {Ingram}}, \bibinfo
  {author} {\bibfnamefont {S.}~\bibnamefont {Freear}}, \ and\ \bibinfo {author}
  {\bibfnamefont {J.~R.}\ \bibnamefont {McLaughlan}},\ }\bibfield  {title}
  {\enquote {\bibinfo {title} {Optimising gold nanorods for photoacoustic
  imaging in vitro},}\ }\href@noop {} {\bibfield  {journal} {\bibinfo
  {journal} {Nanoscale Advances}\ }\textbf {\bibinfo {volume} {1}},\ \bibinfo
  {pages} {1472--1481} (\bibinfo {year} {2019})}\BibitemShut {NoStop}%
\bibitem [{\citenamefont {Plech}\ \emph {et~al.}(2004)\citenamefont {Plech},
  \citenamefont {Kotaidis}, \citenamefont {Gr{\'e}sillon}, \citenamefont
  {Dahmen},\ and\ \citenamefont {Von~Plessen}}]{plech2004laser}%
  \BibitemOpen
  \bibfield  {author} {\bibinfo {author} {\bibfnamefont {A.}~\bibnamefont
  {Plech}}, \bibinfo {author} {\bibfnamefont {V.}~\bibnamefont {Kotaidis}},
  \bibinfo {author} {\bibfnamefont {S.}~\bibnamefont {Gr{\'e}sillon}}, \bibinfo
  {author} {\bibfnamefont {C.}~\bibnamefont {Dahmen}}, \ and\ \bibinfo {author}
  {\bibfnamefont {G.}~\bibnamefont {Von~Plessen}},\ }\bibfield  {title}
  {\enquote {\bibinfo {title} {Laser-induced heating and melting of gold
  nanoparticles studied by time-resolved x-ray scattering},}\ }\href@noop {}
  {\bibfield  {journal} {\bibinfo  {journal} {Physical Review B}\ }\textbf
  {\bibinfo {volume} {70}},\ \bibinfo {pages} {195423} (\bibinfo {year}
  {2004})}\BibitemShut {NoStop}%
\bibitem [{\citenamefont {Stoll}\ \emph {et~al.}(2015)\citenamefont {Stoll},
  \citenamefont {Maioli}, \citenamefont {Crut}, \citenamefont {Rodal-Cedeira},
  \citenamefont {Pastoriza-Santos}, \citenamefont {Vall{\'e}e},\ and\
  \citenamefont {Del~Fatti}}]{stoll2015time}%
  \BibitemOpen
  \bibfield  {author} {\bibinfo {author} {\bibfnamefont {T.}~\bibnamefont
  {Stoll}}, \bibinfo {author} {\bibfnamefont {P.}~\bibnamefont {Maioli}},
  \bibinfo {author} {\bibfnamefont {A.}~\bibnamefont {Crut}}, \bibinfo {author}
  {\bibfnamefont {S.}~\bibnamefont {Rodal-Cedeira}}, \bibinfo {author}
  {\bibfnamefont {I.}~\bibnamefont {Pastoriza-Santos}}, \bibinfo {author}
  {\bibfnamefont {F.}~\bibnamefont {Vall{\'e}e}}, \ and\ \bibinfo {author}
  {\bibfnamefont {N.}~\bibnamefont {Del~Fatti}},\ }\bibfield  {title} {\enquote
  {\bibinfo {title} {Time-resolved investigations of the cooling dynamics of
  metal nanoparticles: impact of environment},}\ }\href@noop {} {\bibfield
  {journal} {\bibinfo  {journal} {The Journal of Physical Chemistry C}\
  }\textbf {\bibinfo {volume} {119}},\ \bibinfo {pages} {12757--12764}
  (\bibinfo {year} {2015})}\BibitemShut {NoStop}%
\bibitem [{\citenamefont {Wilson}\ \emph {et~al.}(2022)\citenamefont {Wilson},
  \citenamefont {Nielsen}, \citenamefont {Randrianalisoa},\ and\ \citenamefont
  {Qin}}]{wilson2022curvature}%
  \BibitemOpen
  \bibfield  {author} {\bibinfo {author} {\bibfnamefont {B.~A.}\ \bibnamefont
  {Wilson}}, \bibinfo {author} {\bibfnamefont {S.~O.}\ \bibnamefont {Nielsen}},
  \bibinfo {author} {\bibfnamefont {J.~H.}\ \bibnamefont {Randrianalisoa}}, \
  and\ \bibinfo {author} {\bibfnamefont {Z.}~\bibnamefont {Qin}},\ }\bibfield
  {title} {\enquote {\bibinfo {title} {Curvature and temperature-dependent
  thermal interface conductance between nanoscale gold and water},}\
  }\href@noop {} {\bibfield  {journal} {\bibinfo  {journal} {The Journal of
  Chemical Physics}\ }\textbf {\bibinfo {volume} {157}},\ \bibinfo {pages}
  {054703} (\bibinfo {year} {2022})}\BibitemShut {NoStop}%
\bibitem [{\citenamefont {Herrero}, \citenamefont {Joly},\ and\ \citenamefont
  {Merabia}(2022)}]{herrero2022ultra}%
  \BibitemOpen
  \bibfield  {author} {\bibinfo {author} {\bibfnamefont {C.}~\bibnamefont
  {Herrero}}, \bibinfo {author} {\bibfnamefont {L.}~\bibnamefont {Joly}}, \
  and\ \bibinfo {author} {\bibfnamefont {S.}~\bibnamefont {Merabia}},\
  }\bibfield  {title} {\enquote {\bibinfo {title} {Ultra-high liquid--solid
  thermal resistance using nanostructured gold surfaces coated with
  graphene},}\ }\href@noop {} {\bibfield  {journal} {\bibinfo  {journal}
  {Applied Physics Letters}\ }\textbf {\bibinfo {volume} {120}},\ \bibinfo
  {pages} {171601} (\bibinfo {year} {2022})}\BibitemShut {NoStop}%
\bibitem [{Note1()}]{Note1}%
  \BibitemOpen
  \bibinfo {note} {Due to the intricate relaxation dynamics, $\tau _{th}$
  escapes a formal definition. A commonly adopted operative approach is to
  define it as the time necessary for the GNC temperature increase, triggered
  by a $\delta $-like excitation source, to fall to 1/e of its maximum\cite
  {diego2022Ultrafast}.}\BibitemShut {Stop}%
\bibitem [{\citenamefont {Han}, \citenamefont {M\'erabia},\ and\ \citenamefont
  {M\"uller-Plathe}(2017)}]{han2017thermal}%
  \BibitemOpen
  \bibfield  {author} {\bibinfo {author} {\bibfnamefont {H.}~\bibnamefont
  {Han}}, \bibinfo {author} {\bibfnamefont {S.}~\bibnamefont {M\'erabia}}, \
  and\ \bibinfo {author} {\bibfnamefont {F.}~\bibnamefont {M\"uller-Plathe}},\
  }\bibfield  {title} {\enquote {\bibinfo {title} {Thermal transport at
  solid--liquid interfaces: High pressure facilitates heat flow through
  nonlocal liquid structuring},}\ }\href@noop {} {\bibfield  {journal}
  {\bibinfo  {journal} {The Journal of Physical Chemistry Letters}\ }\textbf
  {\bibinfo {volume} {8}},\ \bibinfo {pages} {1946--1951} (\bibinfo {year}
  {2017})}\BibitemShut {NoStop}%
\bibitem [{\citenamefont {Pham}, \citenamefont {Barisik},\ and\ \citenamefont
  {Kim}(2013)}]{pham2013pressure}%
  \BibitemOpen
  \bibfield  {author} {\bibinfo {author} {\bibfnamefont {A.}~\bibnamefont
  {Pham}}, \bibinfo {author} {\bibfnamefont {M.}~\bibnamefont {Barisik}}, \
  and\ \bibinfo {author} {\bibfnamefont {B.}~\bibnamefont {Kim}},\ }\bibfield
  {title} {\enquote {\bibinfo {title} {Pressure dependence of kapitza
  resistance at gold/water and silicon/water interfaces},}\ }\href@noop {}
  {\bibfield  {journal} {\bibinfo  {journal} {The Journal of Chemical Physics}\
  }\textbf {\bibinfo {volume} {139}},\ \bibinfo {pages} {244702} (\bibinfo
  {year} {2013})}\BibitemShut {NoStop}%
\bibitem [{\citenamefont {Barisik}\ and\ \citenamefont
  {Beskok}(2014)}]{barisik2014temperature}%
  \BibitemOpen
  \bibfield  {author} {\bibinfo {author} {\bibfnamefont {M.}~\bibnamefont
  {Barisik}}\ and\ \bibinfo {author} {\bibfnamefont {A.}~\bibnamefont
  {Beskok}},\ }\bibfield  {title} {\enquote {\bibinfo {title} {Temperature
  dependence of thermal resistance at the water/silicon interface},}\
  }\href@noop {} {\bibfield  {journal} {\bibinfo  {journal} {International
  Journal of Thermal Sciences}\ }\textbf {\bibinfo {volume} {77}},\ \bibinfo
  {pages} {47--54} (\bibinfo {year} {2014})}\BibitemShut {NoStop}%
\bibitem [{\citenamefont {Vera}\ and\ \citenamefont
  {Bayazitoglu}(2015)}]{vera2015temperature}%
  \BibitemOpen
  \bibfield  {author} {\bibinfo {author} {\bibfnamefont {J.}~\bibnamefont
  {Vera}}\ and\ \bibinfo {author} {\bibfnamefont {Y.}~\bibnamefont
  {Bayazitoglu}},\ }\bibfield  {title} {\enquote {\bibinfo {title} {Temperature
  and heat flux dependence of thermal resistance of water/metal nanoparticle
  interfaces at sub-boiling temperatures},}\ }\href@noop {} {\bibfield
  {journal} {\bibinfo  {journal} {International Journal of Heat and Mass
  Transfer}\ }\textbf {\bibinfo {volume} {86}},\ \bibinfo {pages} {433--442}
  (\bibinfo {year} {2015})}\BibitemShut {NoStop}%
\bibitem [{\citenamefont {Banfi}\ \emph {et~al.}(2012)\citenamefont {Banfi},
  \citenamefont {Juv{\'e}}, \citenamefont {Nardi}, \citenamefont {Dal~Conte},
  \citenamefont {Giannetti}, \citenamefont {Ferrini}, \citenamefont
  {Del~Fatti},\ and\ \citenamefont {Vall{\'e}e}}]{banfi2012temperature}%
  \BibitemOpen
  \bibfield  {author} {\bibinfo {author} {\bibfnamefont {F.}~\bibnamefont
  {Banfi}}, \bibinfo {author} {\bibfnamefont {V.}~\bibnamefont {Juv{\'e}}},
  \bibinfo {author} {\bibfnamefont {D.}~\bibnamefont {Nardi}}, \bibinfo
  {author} {\bibfnamefont {S.}~\bibnamefont {Dal~Conte}}, \bibinfo {author}
  {\bibfnamefont {C.}~\bibnamefont {Giannetti}}, \bibinfo {author}
  {\bibfnamefont {G.}~\bibnamefont {Ferrini}}, \bibinfo {author} {\bibfnamefont
  {N.}~\bibnamefont {Del~Fatti}}, \ and\ \bibinfo {author} {\bibfnamefont
  {F.}~\bibnamefont {Vall{\'e}e}},\ }\bibfield  {title} {\enquote {\bibinfo
  {title} {Temperature dependence of the thermal boundary resistivity of
  glass-embedded metal nanoparticles},}\ }\href@noop {} {\bibfield  {journal}
  {\bibinfo  {journal} {Applied Physics Letters}\ }\textbf {\bibinfo {volume}
  {100}},\ \bibinfo {pages} {011902} (\bibinfo {year} {2012})}\BibitemShut
  {NoStop}%
\bibitem [{\citenamefont {Wu}\ \emph {et~al.}(2016)\citenamefont {Wu},
  \citenamefont {Ni}, \citenamefont {Zhu}, \citenamefont {Burrows},
  \citenamefont {Murphy}, \citenamefont {Dumitrica},\ and\ \citenamefont
  {Wang}}]{wu2016thermal}%
  \BibitemOpen
  \bibfield  {author} {\bibinfo {author} {\bibfnamefont {X.}~\bibnamefont
  {Wu}}, \bibinfo {author} {\bibfnamefont {Y.}~\bibnamefont {Ni}}, \bibinfo
  {author} {\bibfnamefont {J.}~\bibnamefont {Zhu}}, \bibinfo {author}
  {\bibfnamefont {N.~D.}\ \bibnamefont {Burrows}}, \bibinfo {author}
  {\bibfnamefont {C.~J.}\ \bibnamefont {Murphy}}, \bibinfo {author}
  {\bibfnamefont {T.}~\bibnamefont {Dumitrica}}, \ and\ \bibinfo {author}
  {\bibfnamefont {X.}~\bibnamefont {Wang}},\ }\bibfield  {title} {\enquote
  {\bibinfo {title} {Thermal transport across surfactant layers on gold
  nanorods in aqueous solution},}\ }\href@noop {} {\bibfield  {journal}
  {\bibinfo  {journal} {ACS Applied Materials \& Interfaces}\ }\textbf
  {\bibinfo {volume} {8}},\ \bibinfo {pages} {10581--10589} (\bibinfo {year}
  {2016})}\BibitemShut {NoStop}%
\bibitem [{\citenamefont {Caplan}, \citenamefont {Giri},\ and\ \citenamefont
  {Hopkins}(2014)}]{caplan2014analytical}%
  \BibitemOpen
  \bibfield  {author} {\bibinfo {author} {\bibfnamefont {M.~E.}\ \bibnamefont
  {Caplan}}, \bibinfo {author} {\bibfnamefont {A.}~\bibnamefont {Giri}}, \ and\
  \bibinfo {author} {\bibfnamefont {P.~E.}\ \bibnamefont {Hopkins}},\
  }\bibfield  {title} {\enquote {\bibinfo {title} {Analytical model for the
  effects of wetting on thermal boundary conductance across solid/classical
  liquid interfaces},}\ }\href@noop {} {\bibfield  {journal} {\bibinfo
  {journal} {The Journal of Chemical Physics}\ }\textbf {\bibinfo {volume}
  {140}},\ \bibinfo {pages} {154701} (\bibinfo {year} {2014})}\BibitemShut
  {NoStop}%
\bibitem [{\citenamefont {Ramos-Alvarado}, \citenamefont {Kumar},\ and\
  \citenamefont {Peterson}(2016)}]{ramos2016solid}%
  \BibitemOpen
  \bibfield  {author} {\bibinfo {author} {\bibfnamefont {B.}~\bibnamefont
  {Ramos-Alvarado}}, \bibinfo {author} {\bibfnamefont {S.}~\bibnamefont
  {Kumar}}, \ and\ \bibinfo {author} {\bibfnamefont {G.}~\bibnamefont
  {Peterson}},\ }\bibfield  {title} {\enquote {\bibinfo {title} {Solid--liquid
  thermal transport and its relationship with wettability and the interfacial
  liquid structure},}\ }\href@noop {} {\bibfield  {journal} {\bibinfo
  {journal} {The Journal of Physical Chemistry Letters}\ }\textbf {\bibinfo
  {volume} {7}},\ \bibinfo {pages} {3497--3501} (\bibinfo {year}
  {2016})}\BibitemShut {NoStop}%
\bibitem [{\citenamefont {Kim}, \citenamefont {Beskok},\ and\ \citenamefont
  {Cagin}(2008)}]{kim2008molecular}%
  \BibitemOpen
  \bibfield  {author} {\bibinfo {author} {\bibfnamefont {B.~H.}\ \bibnamefont
  {Kim}}, \bibinfo {author} {\bibfnamefont {A.}~\bibnamefont {Beskok}}, \ and\
  \bibinfo {author} {\bibfnamefont {T.}~\bibnamefont {Cagin}},\ }\bibfield
  {title} {\enquote {\bibinfo {title} {Molecular dynamics simulations of
  thermal resistance at the liquid-solid interface},}\ }\href@noop {}
  {\bibfield  {journal} {\bibinfo  {journal} {The Journal of Chemical Physics}\
  }\textbf {\bibinfo {volume} {129}},\ \bibinfo {pages} {174701} (\bibinfo
  {year} {2008})}\BibitemShut {NoStop}%
\bibitem [{\citenamefont {Vo}\ \emph {et~al.}(2015)\citenamefont {Vo},
  \citenamefont {Park}, \citenamefont {Park},\ and\ \citenamefont
  {Kim}}]{vo2015nano}%
  \BibitemOpen
  \bibfield  {author} {\bibinfo {author} {\bibfnamefont {T.~Q.}\ \bibnamefont
  {Vo}}, \bibinfo {author} {\bibfnamefont {B.}~\bibnamefont {Park}}, \bibinfo
  {author} {\bibfnamefont {C.}~\bibnamefont {Park}}, \ and\ \bibinfo {author}
  {\bibfnamefont {B.}~\bibnamefont {Kim}},\ }\bibfield  {title} {\enquote
  {\bibinfo {title} {Nano-scale liquid film sheared between strong wetting
  surfaces: Effects of interface region on the flow},}\ }\href@noop {}
  {\bibfield  {journal} {\bibinfo  {journal} {Journal of Mechanical Science and
  Technology}\ }\textbf {\bibinfo {volume} {29}},\ \bibinfo {pages}
  {1681--1688} (\bibinfo {year} {2015})}\BibitemShut {NoStop}%
\bibitem [{\citenamefont {Barrat}\ and\ \citenamefont
  {Chiaruttini}(2003)}]{barrat2003kapitza}%
  \BibitemOpen
  \bibfield  {author} {\bibinfo {author} {\bibfnamefont {J.-L.}\ \bibnamefont
  {Barrat}}\ and\ \bibinfo {author} {\bibfnamefont {F.}~\bibnamefont
  {Chiaruttini}},\ }\bibfield  {title} {\enquote {\bibinfo {title} {Kapitza
  resistance at the liquid—solid interface},}\ }\href@noop {} {\bibfield
  {journal} {\bibinfo  {journal} {Molecular Physics}\ }\textbf {\bibinfo
  {volume} {101}},\ \bibinfo {pages} {1605--1610} (\bibinfo {year}
  {2003})}\BibitemShut {NoStop}%
\bibitem [{\citenamefont {Hu}\ and\ \citenamefont {Sun}(2012)}]{hu2012effect}%
  \BibitemOpen
  \bibfield  {author} {\bibinfo {author} {\bibfnamefont {H.}~\bibnamefont
  {Hu}}\ and\ \bibinfo {author} {\bibfnamefont {Y.}~\bibnamefont {Sun}},\
  }\bibfield  {title} {\enquote {\bibinfo {title} {Effect of nanopatterns on
  kapitza resistance at a water-gold interface during boiling: A molecular
  dynamics study},}\ }\href@noop {} {\bibfield  {journal} {\bibinfo  {journal}
  {Journal of Applied Physics}\ }\textbf {\bibinfo {volume} {112}},\ \bibinfo
  {pages} {053508} (\bibinfo {year} {2012})}\BibitemShut {NoStop}%
\bibitem [{\citenamefont {Wang}\ and\ \citenamefont
  {Keblinski}(2011)}]{wang2011role}%
  \BibitemOpen
  \bibfield  {author} {\bibinfo {author} {\bibfnamefont {Y.}~\bibnamefont
  {Wang}}\ and\ \bibinfo {author} {\bibfnamefont {P.}~\bibnamefont
  {Keblinski}},\ }\bibfield  {title} {\enquote {\bibinfo {title} {Role of
  wetting and nanoscale roughness on thermal conductance at liquid-solid
  interface},}\ }\href@noop {} {\bibfield  {journal} {\bibinfo  {journal}
  {Applied Physics Letters}\ }\textbf {\bibinfo {volume} {99}},\ \bibinfo
  {pages} {073112} (\bibinfo {year} {2011})}\BibitemShut {NoStop}%
\bibitem [{\citenamefont {Hasan}, \citenamefont {Vo},\ and\ \citenamefont
  {Kim}(2019)}]{hasan2019manipulating}%
  \BibitemOpen
  \bibfield  {author} {\bibinfo {author} {\bibfnamefont {M.~R.}\ \bibnamefont
  {Hasan}}, \bibinfo {author} {\bibfnamefont {T.~Q.}\ \bibnamefont {Vo}}, \
  and\ \bibinfo {author} {\bibfnamefont {B.}~\bibnamefont {Kim}},\ }\bibfield
  {title} {\enquote {\bibinfo {title} {Manipulating thermal resistance at the
  solid--fluid interface through monolayer deposition},}\ }\href@noop {}
  {\bibfield  {journal} {\bibinfo  {journal} {RSC advances}\ }\textbf {\bibinfo
  {volume} {9}},\ \bibinfo {pages} {4948--4956} (\bibinfo {year}
  {2019})}\BibitemShut {NoStop}%
\bibitem [{\citenamefont {Yenigun}\ and\ \citenamefont
  {Barisik}(2019)}]{yenigun2019effect}%
  \BibitemOpen
  \bibfield  {author} {\bibinfo {author} {\bibfnamefont {O.}~\bibnamefont
  {Yenigun}}\ and\ \bibinfo {author} {\bibfnamefont {M.}~\bibnamefont
  {Barisik}},\ }\bibfield  {title} {\enquote {\bibinfo {title} {Effect of
  nano-film thickness on thermal resistance at water/silicon interface},}\
  }\href@noop {} {\bibfield  {journal} {\bibinfo  {journal} {International
  Journal of Heat and Mass Transfer}\ }\textbf {\bibinfo {volume} {134}},\
  \bibinfo {pages} {634--640} (\bibinfo {year} {2019})}\BibitemShut {NoStop}%
\bibitem [{\citenamefont {Cao}\ \emph {et~al.}(2018)\citenamefont {Cao},
  \citenamefont {Zou}, \citenamefont {Hu},\ and\ \citenamefont
  {Cao}}]{cao2018enhanced}%
  \BibitemOpen
  \bibfield  {author} {\bibinfo {author} {\bibfnamefont {B.-Y.}\ \bibnamefont
  {Cao}}, \bibinfo {author} {\bibfnamefont {J.-H.}\ \bibnamefont {Zou}},
  \bibinfo {author} {\bibfnamefont {G.-J.}\ \bibnamefont {Hu}}, \ and\ \bibinfo
  {author} {\bibfnamefont {G.-X.}\ \bibnamefont {Cao}},\ }\bibfield  {title}
  {\enquote {\bibinfo {title} {Enhanced thermal transport across multilayer
  graphene and water by interlayer functionalization},}\ }\href@noop {}
  {\bibfield  {journal} {\bibinfo  {journal} {Applied Physics Letters}\
  }\textbf {\bibinfo {volume} {112}},\ \bibinfo {pages} {041603} (\bibinfo
  {year} {2018})}\BibitemShut {NoStop}%
\bibitem [{\citenamefont {Tian}, \citenamefont {Marconnet},\ and\ \citenamefont
  {Chen}(2015)}]{tian2015enhancing}%
  \BibitemOpen
  \bibfield  {author} {\bibinfo {author} {\bibfnamefont {Z.}~\bibnamefont
  {Tian}}, \bibinfo {author} {\bibfnamefont {A.}~\bibnamefont {Marconnet}}, \
  and\ \bibinfo {author} {\bibfnamefont {G.}~\bibnamefont {Chen}},\ }\bibfield
  {title} {\enquote {\bibinfo {title} {Enhancing solid-liquid interface thermal
  transport using self-assembled monolayers},}\ }\href@noop {} {\bibfield
  {journal} {\bibinfo  {journal} {Applied Physics Letters}\ }\textbf {\bibinfo
  {volume} {106}},\ \bibinfo {pages} {211602} (\bibinfo {year}
  {2015})}\BibitemShut {NoStop}%
\bibitem [{Note2()}]{Note2}%
  \BibitemOpen
  \bibinfo {note} {Formally, the GNC cooling is mono-exponential with a decay
  given by this formula, only for a Biot number Bi$\ll $1 and for isothermal
  water \cite {banfi2010ab}, i.e. the cooling is limited by the interfacial
  heat transfer only.}\BibitemShut {Stop}%
\bibitem [{Note3()}]{Note3}%
  \BibitemOpen
  \bibinfo {note} {As an extreme case scenario, if we were to nullify the TBR
  we would have $\tau _{TBR}$=0. In such a situation, the GNC thermal dynamics
  is entirely ruled by the GNC and proximal water thermal inertia.}\BibitemShut
  {Stop}%
\bibitem [{Note4()}]{Note4}%
  \BibitemOpen
  \bibinfo {note} {Note however that, even for the 10 ps light pulse case, the
  mechanophone effect for $\tau _{TBR}/\tau _L$=10$^{2}$ is lower than the
  thermophone effect for $\tau _{TBR}/\tau _L$=10$^{-2}$. At room temperature,
  water thermal expansion coefficient is $\sim $ 5 times higher than the gold's
  one. Water's expansion is then more efficient than gold's, leading to the
  maximum of the thermophone effect to exceed that of the mechanophone one. We
  tested that, using the same thermal expansion coefficients for both the GNC
  and water results in a maximum of the mechanophone effect (occurring at $\tau
  _{TBR}/\tau _L$=10$^{-2}$) to be slightly higher than the maximum thermophone
  effect (occurring at $\tau _{Th}/\tau _L$=10$^{2}$).}\BibitemShut {Stop}%
\bibitem [{\citenamefont {Bertolotti}\ and\ \citenamefont
  {Li~Voti}(2020)}]{bertolotti2020note}%
  \BibitemOpen
  \bibfield  {author} {\bibinfo {author} {\bibfnamefont {M.}~\bibnamefont
  {Bertolotti}}\ and\ \bibinfo {author} {\bibfnamefont {R.}~\bibnamefont
  {Li~Voti}},\ }\bibfield  {title} {\enquote {\bibinfo {title} {A note on the
  history of photoacoustic, thermal lensing, and photothermal deflection
  techniques},}\ }\href@noop {} {\bibfield  {journal} {\bibinfo  {journal}
  {Journal of Applied Physics}\ }\textbf {\bibinfo {volume} {128}},\ \bibinfo
  {pages} {230901} (\bibinfo {year} {2020})}\BibitemShut {NoStop}%
\bibitem [{\citenamefont {Gandolfi}\ \emph {et~al.}(2022)\citenamefont
  {Gandolfi}, \citenamefont {Peli}, \citenamefont {Diego}, \citenamefont
  {Danesi}, \citenamefont {Giannetti}, \citenamefont {Alessandri},
  \citenamefont {Zannier}, \citenamefont {Demontis}, \citenamefont {Rocci},
  \citenamefont {Beltram}, \citenamefont {Sorba}, \citenamefont {Roddaro},
  \citenamefont {Rossella},\ and\ \citenamefont
  {Banfi}}]{gandolfi2022ultrafast}%
  \BibitemOpen
  \bibfield  {author} {\bibinfo {author} {\bibfnamefont {M.}~\bibnamefont
  {Gandolfi}}, \bibinfo {author} {\bibfnamefont {S.}~\bibnamefont {Peli}},
  \bibinfo {author} {\bibfnamefont {M.}~\bibnamefont {Diego}}, \bibinfo
  {author} {\bibfnamefont {S.}~\bibnamefont {Danesi}}, \bibinfo {author}
  {\bibfnamefont {C.}~\bibnamefont {Giannetti}}, \bibinfo {author}
  {\bibfnamefont {I.}~\bibnamefont {Alessandri}}, \bibinfo {author}
  {\bibfnamefont {V.}~\bibnamefont {Zannier}}, \bibinfo {author} {\bibfnamefont
  {V.}~\bibnamefont {Demontis}}, \bibinfo {author} {\bibfnamefont
  {M.}~\bibnamefont {Rocci}}, \bibinfo {author} {\bibfnamefont
  {F.}~\bibnamefont {Beltram}}, \bibinfo {author} {\bibfnamefont
  {L.}~\bibnamefont {Sorba}}, \bibinfo {author} {\bibfnamefont
  {S.}~\bibnamefont {Roddaro}}, \bibinfo {author} {\bibfnamefont
  {F.}~\bibnamefont {Rossella}}, \ and\ \bibinfo {author} {\bibfnamefont
  {F.}~\bibnamefont {Banfi}},\ }\bibfield  {title} {\enquote {\bibinfo {title}
  {Ultrafast photoacoustic nanometrology of inas nanowires mechanical
  properties},}\ }\href@noop {} {\bibfield  {journal} {\bibinfo  {journal} {The
  Journal of Physical Chemistry C}\ }\textbf {\bibinfo {volume} {126}},\
  \bibinfo {pages} {6361--6372} (\bibinfo {year} {2022})}\BibitemShut {NoStop}%
\bibitem [{\citenamefont {Banfi}\ \emph {et~al.}(2010)\citenamefont {Banfi},
  \citenamefont {Pressacco}, \citenamefont {Revaz}, \citenamefont {Giannetti},
  \citenamefont {Nardi}, \citenamefont {Ferrini},\ and\ \citenamefont
  {Parmigiani}}]{banfi2010ab}%
  \BibitemOpen
  \bibfield  {author} {\bibinfo {author} {\bibfnamefont {F.}~\bibnamefont
  {Banfi}}, \bibinfo {author} {\bibfnamefont {F.}~\bibnamefont {Pressacco}},
  \bibinfo {author} {\bibfnamefont {B.}~\bibnamefont {Revaz}}, \bibinfo
  {author} {\bibfnamefont {C.}~\bibnamefont {Giannetti}}, \bibinfo {author}
  {\bibfnamefont {D.}~\bibnamefont {Nardi}}, \bibinfo {author} {\bibfnamefont
  {G.}~\bibnamefont {Ferrini}}, \ and\ \bibinfo {author} {\bibfnamefont
  {F.}~\bibnamefont {Parmigiani}},\ }\bibfield  {title} {\enquote {\bibinfo
  {title} {Ab initio thermodynamics calculation of all-optical time-resolved
  calorimetry of nanosize systems: Evidence of nanosecond decoupling of
  electron and phonon temperatures},}\ }\href@noop {} {\bibfield  {journal}
  {\bibinfo  {journal} {Physical Review B}\ }\textbf {\bibinfo {volume} {81}},\
  \bibinfo {pages} {155426} (\bibinfo {year} {2010})}\BibitemShut {NoStop}%
\end{thebibliography}%

\end{document}